%% file: main.tex
\documentclass{IEEE-Con-Sys-mag}

\input{header}

\usepackage[
    backend=biber,
    sorting=none,
    style=ieee,
    citestyle=numeric-comp,
    doi=false,
    url=false,
    isbn=false,
    eprint=false
]{biblatex}

\AtEveryBibitem{%
  \clearfield{issn}%
}
\jname{Working paper}

\renewcommand{\thesequation}{S\arabic{sequation}}
\renewcommand{\thesfigure}{S\arabic{sfigure}}

\begin{document}

\title{Welfarist Control Design\stitle{How to fulfill the societal mandate in multi-agent control?}} 

\author{Sophie Hall, Kai Zhang, Ilia Shilov, Heinrich H. Nax, Saverio Bolognani}

\maketitle

\input{sections/00_introduction}

\section{Short Tutorial on Welfarism}

\input{sections/01_tutorial}

\section{Dynamic Welfarist Control}
\input{sections/02_dynamic_welfarism}

\FloatBarrier

\section{Conclusion and Outlook}

\input{sections/03_outlook_conclusion}

\section*{References}
\printbibliography[heading=none]

\end{document}

%% file: header.tex
\usepackage{tikz}
\usetikzlibrary{patterns}
\usepackage{pgfplots}
\pgfplotsset{compat=1.18}
\usepgfplotslibrary{groupplots}
\usetikzlibrary{shapes.geometric, calc, backgrounds}
\usetikzlibrary{arrows.meta, fit}

\usepackage[outline]{contour}
\contourlength{1pt}

\usepackage{booktabs}
\usepackage{makecell}
\usepackage{amsmath}
\usepackage{amssymb}
\usepackage{standalone}

\usepackage[dvipsnames]{xcolor}
\usepackage{dsfont}
\usepackage{wrapfig}

\usepackage{placeins}

\DeclareMathOperator{\subjectto}{subject\ to}
\DeclareMathOperator*{\argmin}{argmin}

\input{figures/figstyle}

\usepackage{comment}

\makeatletter
\def\paragraph{\@startsection{paragraph}{4}{\z@}
{-6pt \@plus-2pt \@minus-2pt}{.001pt}
{\normalfont\normalsize\bfseries\itshape}}
\makeatother

\newenvironment{axiomp}[2]
  {\par\addvspace{6pt plus 1pt minus 1pt}%
   \noindent\textbf{Axiom (#1) #2 }\\}
  {\par\addvspace{6pt plus 1pt minus 1pt}}

\makeatletter
\renewcommand{\sdbarfig}[2]{%
  \begingroup
  \refstepcounter{sfigure}%
  \bigskip
  {\centering
  #1\\}%
  {\noindent
  {\figcapnumfont FIGURE~\thesfigure\space}%
  {\figcapfont #2\par}%
  }
  \bigskip
  \endgroup
}
\makeatother

\newcommand{\startsidebarfig}{%
  \refstepcounter{sfigure}%
  \let\oldthefigure\thefigure
  \renewcommand{\thefigure}{S\arabic{sfigure}}%
}

\newcommand{\stopsidebarfig}{%
  \addtocounter{figure}{-1}%
  \let\thefigure\oldthefigure
}

%% file: figures/figstyle.tex
\definecolor{CBblue}{RGB}{68,119,170}
\definecolor{CBcyan}{RGB}{102,204,238}
\definecolor{CBgreen}{RGB}{34,136,51}
\definecolor{CByellow}{RGB}{204,187,68}
\definecolor{CBred}{RGB}{238,102,119}
\definecolor{CBpurple}{RGB}{170,51,119}
\definecolor{CBgrey}{RGB}{187,187,187}

\definecolor{CBpaleblue}{RGB}{187,204,238}
\definecolor{CBpalecyan}{RGB}{204,238,255}
\definecolor{CBpalegreen}{RGB}{204,221,170}
\definecolor{CBpaleyellow}{RGB}{238,238,187}
\definecolor{CBpalered}{RGB}{255,204,204}
\definecolor{CBpalegrey}{RGB}{221,221,221}

\definecolor{CBdarkblue}{RGB}{34,34,85}
\definecolor{CBdarkcyan}{RGB}{34,85,85}
\definecolor{CBdarkgreen}{RGB}{34,85,34}
\definecolor{CBdarkyellow}{RGB}{102,102,51}
\definecolor{CBdarkred}{RGB}{102,51,51}
\definecolor{CBdarkgrey}{RGB}{85,85,85}

\colorlet{CBcolor1}{CBblue}
\colorlet{CBcolor2}{CBpurple}
\colorlet{CBcolor3}{CBgreen}
\colorlet{CBcolor4}{CByellow}
\colorlet{CBneutral}{CBdarkgrey}  %
\colorlet{CBgridline}{black!12}

\newlength{\figmainwidth}    
\newlength{\figsidebarwidth} 
\newlength{\figfullwidth}    

\providecommand{\figfont}{\fontsize{6.5}{7.8}\selectfont}

\tikzset{
    csm data/.style={line width=1pt},
    csm aux/.style={line width=0.6pt},
    csm emph/.style={line width=1.6pt},
    csm arrow/.style={-{Stealth[length=4pt,width=3pt]}, line width=0.6pt},
    csm diagram/.style={
        font=\figfont,
        line width=0.6pt,
        >={Stealth[length=4pt,width=3pt]},
    },
}

\pgfplotsset{
    csm/.style={
        axis lines=left,
        axis line style={line width=0.4pt, black, -},
        tick style={line width=0.4pt, black},
        major tick length=2.5pt,
        minor tick length=1.5pt,
        grid=major,
        grid style={line width=0.3pt, CBgridline},
        tick label style={font=\figfont},
        label style={font=\figfont},
        title style={font=\figfont\bfseries, yshift=-2pt},
        every axis plot/.append style={csm data},
        mark size=1.8pt,
        every node/.append style={font=\figfont},
        legend style={
            font=\figfont,
            draw=none,
            fill=none,
            inner sep=1pt,
            /tikz/every even column/.append style={column sep=6pt},
        },
        legend cell align=left,
    },
    csm bars/.style={
        csm,
        ybar stacked,
        bar width=7pt,
        every axis plot/.append style={line width=0.4pt},
        grid=none,
        ymajorgrids,
    },
}

%% file: sections/00_introduction.tex
\begin{figure}[t]
\input{sections/00_introduction/02_summary}
\end{figure}

\input{sections/00_introduction/01_introduction_text}

\input{sections/00_introduction/03_curtailment_example_1}

%% file: sections/00_introduction/02_summary.tex
\begin{summary}
\summaryinitial{A}t the core of most socio-technical systems lies a scarce resource that is allocated among agents: highway lanes, public transit, road space, water rights, energy access, grid capacity, user attention, pollution rights, etc. 
With further automation of the underlying allocation processes, control engineers are increasingly tasked with making decisive assumptions regarding what society wants.
In practice to date, design choices are largely driven by industry norms and conventions rather than a result of conscientiously responsible and ethical design. 
In this paper, we look at tools available to control engineers to design systems in a more principled manner in order to match the societal mandate.
We consider three control design paradigms: online feedback optimization, control of Markov decision processes, and model predictive control. 
Beginning with aggregating individual agents' preferences into control design objectives, subsequently ensuring and certifying the fulfillment of those specifications, we argue that the feedback nature of control systems enables appropriate allocation of the shared resources in ways hitherto unparalleled.
\end{summary}

%% file: sections/00_introduction/01_introduction_text.tex
\begin{pullquote}
Similarly to how control theorists have established standards to certify stability, robustness, and performance of a control system, the industry also urgently needs a framework to certify the fulfillment of the societal mandate for fair (efficient and equitable) use of shared resources.
\end{pullquote}

\chapterinitial{C}entral challenges for a society are distributing goods and regulating access to services that meet primary needs for its population: water, food, transportation, energy, communication, and healthcare. Implicit in all of them is a collective agreement that society has some kind of governance mandate to ensure universal access, fairness, safety, sustainability, and accountability in the production and delivery.

In many modern societies, allocating and servicing is increasingly automated, meaning that access is granted dynamically and resources are distributed in real time. This creates challenges and, more importantly, opportunities to fulfill the values underpinning the societal mandate more closely and more efficiently than ever before. The core challenge resulting from the scarcity of the resource that needs to be distributed lies in the heterogeneous nature of the recipients, who experience different utility or cost when receiving the same service or good. Ideally, when aligned with the societal mandate, automation can trade off these competing interests fairly or can satisfy them with fewer resources (that is, more sustainably).

Trading off individual interests in a way that matches the societal mandate across alternative allocations is really at the heart of many socio-technical systems. 
In control, this is pursued under a variety of headers including Pareto efficient control (multi-objective optimization)~\cite{gambier2007multi}, cooperative game theory~\cite{fele2017coalitional}, and equity-, equality- and fairness-oriented control~\cite{radunovic2007unified} and programming \cite{Bhadoriya2026-kp,Sundar2025-ec}. Successful examples which have been implemented at scale include proportional fairness across the Internet \cite{KellyMaullooTan1998,chiu1989analysis,mo2000fair}, fair scheduling in cellular base stations \cite{kushner2004convergence}, and slot allocation in air traffic flow management~\cite{vossen2003general}. Many of the resulting control systems are designed for specific domain applications based on particular or ad hoc objectives and constraints, given one particular societal mandate.

\begin{figure}
    \centering
    \includegraphics{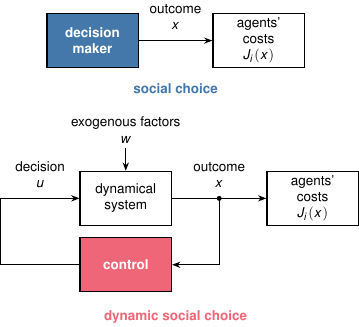}
    \caption{In standard social choice, a welfarist stance is used to select an outcome based on the agents' preferences. In control problems, repeated decisions are taken and affect the outcome through a dynamical system. The goal of \emph{welfarist control design} is to align the \emph{closed-loop behavior} to the preferences of the agents.   }
    \label{fig:dynamic-social-choice}
\end{figure}

To generalize beyond specific instances, and given the growing number of applications affecting human livelihood, control theory requires a principled approach to identify the right trade-offs. 
Evidently, this will be very hard in general, even if the objective is relatively clear, say ``equality.'' 
Equality of what~\cite{Sen1980EqualityOfWhat,Moulin2003}? Fundamental questions and issues immediately arise; ex-ante (equal expected outcomes) versus ex-post (equal realizations), in allocated resources or in cost outcomes, etc. 
Justifying inequality is even harder. Which circumstances and parameters justify unequal treatments, how, when, and to what extent? And which inequalities are tolerated: financial, need-based, structural? And ought they be cross-cut between domains?

\begin{sidebar}{No Fairness Without Welfarism}
Research on fairness has shown that the perceived fairness of the distribution and allocation of a resource among those being served is critical to the societal acceptance of the allocation rule and procedure \cite{gross2007community,almaas2010fairness}.
Not only which outcomes are reached (who gets what) but also how these outcomes are reached procedurally determine whether a solution is endorsed/accepted or not \cite{esaiasson2019reconsidering, hollander2008procedural}.
As the distribution of many resources has been increasingly automated, control theorists have shown growing interest in designing ``fair control.''

Trading off equality and efficiency has been referred to as ``the big trade-off'' \cite{okun2015equality}, and many fairness indices used in control can be understood as cutting that trade-off in a specific way. Any such index is a joint measure that takes the preferences of all affected agents and selects a ``fair'' outcome. In doing so, it inherently compares the benefits and losses of individuals through that joint measure. In practice, the choice of such an index is often guided by rules, values, or principles that the allocation should satisfy according to the engineer designing the automation system. Such indices may embed unintended implicit assumptions and can fail to capture heterogeneity in how agents benefit from resources. 

Fair control often uses the same ingredients that welfarist social choice has established, yet without a principled underpinning. The framework presented in this paper gives a full design pipeline for welfarist control, centered in social choice theory, that makes these choices explicit and shows that a variety of allocations along the spectrum can be recovered and justified from them. Several, but not all, social cost functions with attractive fairness properties turn out to be permissible, depending on the context. 

\sdbarfig{%
\centering
\includegraphics{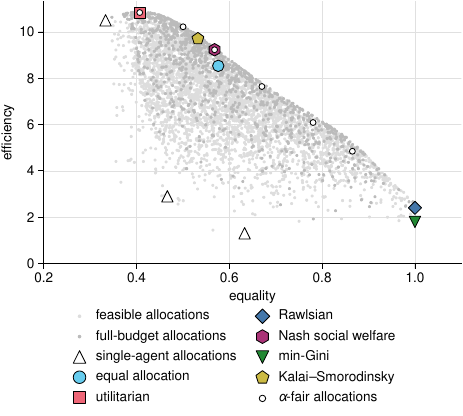}\medskip}%
{Consider allocating resources to a set of agents with heterogeneous utility functions $u_1(x_1)=0.5+10\sqrt{x_1}$, $u_2(x_2)=\ln(1+10x_2)$, $u_3(x_3)=0.8\left(1-e^{-15x_3}\right)$, 
subject to a budget constraint $x_1+x_2+x_3 \le 1$. Possible allocations differ in efficiency (sum of agents' utilities) and equality (e.g., Jain index). Any ``fair'' solution concept implicitly aims at a trade-off between efficiency and equality. It is not a simple design space to navigate:  notice that some points where budget is unused are, at the same time, more efficient and more equal than points where the full budget is distributed.}

\end{sidebar}

The theory of social choice, a foundational theory for welfare economics \cite{bergson1938reformulation, samuelson1947foundations, Arrow1951, Harsanyi1955, Sen1970CCSW}, has pondered over these issues in the abstract for almost 100 years. Dating back to philosophers like Bentham \cite{bentham1789principles} and Mill \cite{mill1863utilitarianism}, and later reframed by Rawls \cite{Rawls1971}, the crucial question has been which overall social welfare function to choose that encodes the trade-offs between individual utilities.
The main lesson has been that the appropriate choice of function is based on two key ingredients: (i) axioms that any allocation rule needs to satisfy, and (ii) information that is available about individual utilities and how to compare them. Jointly, they might uniquely determine the class of permissible social cost functions, modulo plenty of impossibility results.

\begin{figure}[bt]
    \centering
    \includegraphics{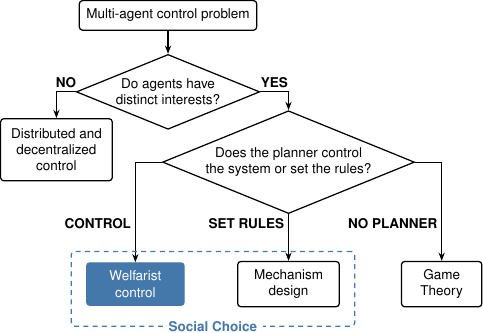}
    \caption{In this paper, we focus on the problem of designing controllers for systems involving multiple agents with distinct interests. Social choice theory is also relevant when agents respond strategically to designed rules, informing the design and evaluation of mechanisms and interaction rules \cite{serrano_theory_2004,paccagnan_utility_2022}, with applications in communication networks \cite{boche_mechanism_2012,chorppath_games_2013}, energy systems \cite{tsaousoglou_mechanism_2021}, mobility \cite{ilia_shilov_invariant_2026}, and socio-technical resource allocation \cite{Elokda2025Vision}. Related work applies the same axiomatic perspective to the efficiency and fairness of strategic equilibria \cite{shilov_invariant_2026, hall_limits_2025,hall2027fairVGNE}.}
    \label{fig:flowchart}
\end{figure}

However, from the viewpoint of control, the results from social choice fly as pie in the sky, as both its impossibility and possibility results are far removed from real-time decision making in socio-technical systems. The original social choice theory was developed as formal philosophy, not as an applied science. Today, perhaps, the time is ripe to make those lessons concrete for applied control problems where automation is effectively already happening and impossibilities are plausibly surmountable, so that the gap can be closed between social choice theory and dynamical systems. That is what we consider welfarist control: a social planner with the authority to control the system takes repeated decisions, under noise and uncertainty, from online measurements, all while adhering to the welfarist axioms and certifiably maximizing social welfare in the long run. Figure \ref{fig:flowchart} places this setting in the broader context of multi-agent control problems.

In this paper, we lay out how to approach this grand challenge, how to tap into the knowledge on axiomatic allocation developed by social choice theory while leveraging the structure of control of systems that are inherently repeated and feature a clear feedback structure (see Figure~\ref{fig:dynamic-social-choice}). We begin by presenting essential results from welfarism \cite{shilov2025welfare} in a short tutorial, laying out fundamental concepts such as social cost functions, interpersonal comparability, welfarist axioms, and summarizing the welfarist design pipeline. In a next step, we develop dynamic welfarism by applying social choice principles to three well-known control settings: feedback optimization, stochastic optimal control, and model predictive control. In doing so, we show that the combination is instructive in both directions. Using social cost functions as control objectives raises new theoretical and computational challenges for static social choice guarantees to hold in dynamic, uncertain, closed-loop settings. On the other hand, dynamical settings offer new opportunities to distribute welfare across agents and over time:
\begin{itemize}
    \item online feedback allows compensating for exogenous factors and defines fairness in terms of outcomes rather than decisions;
    \item welfarist objectives in a stochastic optimal control setting can be formulated to represent ex-ante and ex-post fairness;
    \item predictive control allows using dynamic models to achieve welfarism over time in the presence of constraints.
\end{itemize}
The goal of this manuscript is not to provide final answers but to offer an entry point for curious control theorists to contribute. We focused on (i) making social choice theory accessible through a tutorial, (ii) summarizing initial welfarist control results and providing intuition for how to derive them, (iii) giving an extensive list of open challenges with pointers to relevant literature. 
A principled approach to closed-loop welfarism will allow control engineers to explain and certify real-time decisions in socio-technical systems, an ambitious endeavor that is paramount to ensuring automation's acceptance in society \cite{roadmap2030}.

%% file: sections/00_introduction/03_curtailment_example_1.tex
\begin{sidebar}{Illustrative Example: Fair Power Generation Curtailment}\label{sidebar:fair-curtailment}

\sdbarinitial{C}onsider a group of $N$ prosumers (i.e., users that consume and produce electricity) connected to the same power distribution grid.
At a given time, each prosumer $i$ can generate power up to $\bar p_i \geq 0$.
Technical grid limitations, such as current and voltage limits, prevent the system operator from accepting their full power injection; therefore, their generation needs to be curtailed (in California, about 9\% of the potential output of new utility-scale solar capacity is curtailed \cite{NOVAN2024102930}; about 8\% of potential solar output was curtailed in Texas in 2018 and in California in early 2020 \cite{OShaughnessy2020}).

Grid operators must decide which generators to curtail, guided by multiple (possibly conflicting) criteria: economic cost, sustainability, fair grid access, social equity, and promotion of investments.
These criteria define what can be called \emph{fair generation curtailment} and constitute the \emph{societal mandate} that designers must consider when automating curtailment. 
They are encoded in laws, regulations, and policies, but they can also be the expectation of customers and business entities \cite{MattBolognani2026}.

The current literature addresses this challenge by proposing a variety of rules \cite{gebbran2021fair, lusis2019reducing, liu2020fairness,Borbath2024,Alam2024,Petrou2021}, often without an axiomatic justification. 
We illustrate three examples of curtailment rules based on the following simple formulation.
The grid operator must choose individual power generations $x_i$ such that $0 \leq x_i \leq \bar p_i$ for all $i=1,\ldots,N$, and such that $x \in X$,
the set of feasible power generations based on the grid limits. 
The curtailed amount of prosumer $i$ is therefore 
$\bar p_i - x_i$.

\subsection{Minimum total curtailment} 

One possibility is to minimize the total curtailed energy:
\[
    x^{(1)} \in \arg\min_{x \in X} \sum_{i \in \mathcal N} 
    (\bar p_i - x_i)
\]
This rule is aligned with the sustainability goal of maximizing renewable generation, which is sometimes mandated by law \cite{Schermeyer2018,EEG2017}.
In market-based settings, it also matches the principle that scarce grid capacity should be allocated efficiently.

\subsection{Equal curtailment}
Another possibility is to minimize the largest curtailment:
\[
    x^{(2)} \in \arg\min_{x \in X} \max_{i \in \mathcal N} (\bar p_i - x_i).
\]
It equalizes curtailment in absolute terms, treating the grid as a public good.
Inequitable treatment can worsen existing social disparities \cite{Brockway2021}, and non-discriminatory network access is mandated by grid regulation \cite[Arts. 6, 31]{EU2019_944}.

\subsection{Proportionally fair curtailment}
Another possibility is to maximize the sum of the logarithms of the accepted generations:
\[
    x^{(3)} \in \arg\max_{x \in X} \sum_{i \in \mathcal N} \log x_i .
\]
This is the \emph{proportionally fair} allocation \cite{KellyMaullooTan1998}. It never curtails a prosumer entirely and weighs each curtailment relative to what the prosumer is allowed to generate, which addresses the concern that disproportionate economic risk discourages investment in renewable generation \cite{Cuenca2023}.

\medskip

In these three examples, different societal mandates have been incorporated into the design of a decision rule. 
How can we make this process explicit and rigorous, so that it can be applied, defended, and explained?
And how to generalize that to dynamic settings in which decisions are repeated over time, online measurements are available, and the system is affected by disturbances and uncertainty?

\sdbarfig{%
\includegraphics{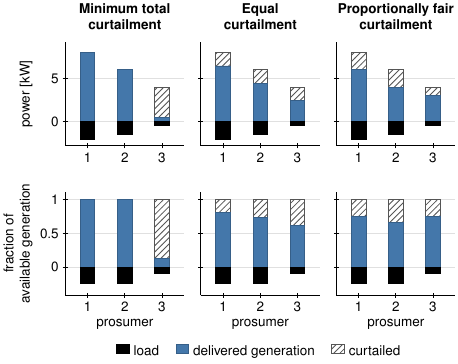}%
}{Example of three prosumers connected to different buses of a power distribution feeder. Different curtailment rules are illustrated in absolute terms (top row) and normalized by each prosumer's available generation (bottom row).
\label{sfig:fair_curtailment_three_rules}}

\end{sidebar}

%% file: sections/01_tutorial.tex
\input{sections/01_tutorial/01_preliminaries}

\input{sections/01_tutorial/02_axioms}

\input{sections/01_tutorial/04_comparability}

\input{sections/01_tutorial/05_summary}

\input{sections/01_tutorial/07_further_reading}

\input{sections/01_tutorial/08_proxy}

%% file: sections/01_tutorial/01_preliminaries.tex
When a system allocates goods or services across multiple agents, a  designer adhering to a welfarist viewpoint cares first and foremost about how those allocations affect the well-being of the agents involved. To do so, the designer must first evaluate, through an individual cost function, what counts as well-being for each agent from each possible outcome, and then base the system-level decision on those individual evaluations. 

Formally, we let $x \in \mathcal{X}$ denote a measurable outcome, be it  a static allocation, a realized service level, a state or a trajectory of the system. Each agent $i$ in a set of agents $\mathcal{N} = \{1, \dots, n\}$ evaluates $x$ through an individual cost function $J_i(\cdot) \in \mathcal{J}$, where $\mathcal{J}$ is the set of all real-valued functions on $\mathcal X$, and $\mathbf{J} = (J_1, \dots, J_n)$ is the associated \emph{cost profile}.

Given a cost profile $\mathbf{J}$ and a feasible set of outcomes $\mathcal{X}$, the welfarist designer is now posed with a question: which outcome would be the ``best'' for the agents? Put differently, how should the feasible outcomes be ranked once the individual evaluations are given? 

We formalize this by introducing, for each cost profile $\mathbf{J}$, a \emph{social preference relation} $\succeq_J$ on $\mathcal{X}$, where $x \succeq_J y$ means that society weakly prefers outcome $x$ to outcome $y$ under profile $\mathbf{J}$. 
We require this relation to be complete and  transitive\footnote{Completeness: for any two outcomes $x,y \in \mathcal{X}$, society can compare them, that is, either $x \succeq_J y$ or $y \succeq_J x$ or both. Transitivity: if $x \succeq_J y$ and $y \succeq_J z$, then $x \succeq_J z$.}, so that it defines a \emph{rational} social ranking over feasible outcomes. 
From $\succeq_J$ we also define strict preference $x \succ_J y$ as $x \succeq_J y$ and not $y \succeq_J x$.
A \emph{social cost functional} is then a rule that assigns such a ranking to every admissible cost profile.

\begin{definition}[Social Cost Functional]
Let $\mathcal{R}$ be the set of complete and transitive binary relations on $\mathcal{X}$. A \emph{Social Cost Functional} (SCFL) is a mapping $F: \mathcal{J}^n \to \mathcal{R}$ that assigns a social preference relation $\succeq_\mathbf{J} = F(\mathbf{J})$ on $\mathcal{X}$ to each cost profile $\mathbf{J} \in \mathcal{J}^{n}$.
\end{definition}

The first task of the welfarist designer is to define properties of the SCFL that are socially desirable.

%% file: sections/01_tutorial/02_axioms.tex
\subsection{Welfarist Axioms}
\label{subsec:axioms}

In the welfarist approach, the desirable properties that any reasonable social ranking should satisfy are formulated axiomatically.
These axioms serve two purposes. 

First, they make the designer's normative requirements explicit. 
In the sidebar ``What breaks without the welfarist axioms'' we provide some examples of what can happen if these axioms are relaxed.

Second, they narrow attention to those SCFLs whose induced social rankings admit a numerical representation by a Social Cost Function (SCF). 

\begin{axiomp}{P}{Pareto Principle}
For any cost profile $\mathbf{J}$ and any $x, y \in \mathcal{X}$: if $J_i(x) < J_i(y)$ for all $i \in \mathcal{N}$, then $x \succ_J y$.
\end{axiomp}

The Pareto Principle is a minimal efficiency requirement: if every agent incurs strictly lower cost under outcome $x$ than under outcome $y$, then society should strictly prefer $x$ to $y$. It rules out choosing an outcome when another feasible outcome is strictly better for every agent.

\begin{sidebar}{What Breaks Without the Welfarist Axioms}
\label{sidebar:what-breaks}

\sdbarinitial{T}he three \emph{welfarist axioms} rule out concrete failure modes in multi-agent decisions. 
We illustrate their importance by discussing what problematic behaviors could happen without those axioms, in a simple case of two agents.

\subsection*{Without Pareto: wasting welfare}

Consider two decisions $x$ and $y$ whose costs for the agents are $\mathbf J(x) = (1,2)$ and $\mathbf J(y) = (3,3)$. Allocation~$x$ Pareto-dominates~$y$: every agent has strictly lower cost under~$x$. Choosing~$y$ over~$x$ would therefore discard an outcome that is better for everyone, which is plainly unsatisfactory. 

Interestingly, a variety of inequality metrics violate this basic axiom. For example, the widely used Gini index \cite{Farris2010} gives $G(\mathbf J(x)) = \tfrac{1}{6}$ and $G( \mathbf J(y)) = 0$, and therefore prefers $y$ over~$x$. The same issue arises with the Jain index, coefficient of variation, and Hoover index (see \cite{chen2023guide} for definitions), which also rank $y$ above~$x$. Inequality metrics may be useful to evaluate the inequality of a system, but optimizing them can produce Pareto-dominated outcomes.

\sdbarfig{%
\includegraphics{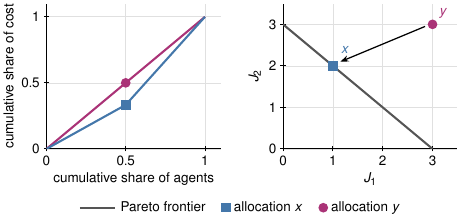}%
}{A Pareto-dominated outcome $y$ can exhibit a better Gini index and therefore be ``preferable'' under that metric, even if no agent prefers it to the alternative.
\label{sfig:gini}}

\subsection*{Without IIA: irrelevant alternatives change the ranking}

The IIA axiom prescribes that the social order between two alternatives $x$ and $y$ depends only on the agents' costs evaluated at $x$ and $y$.

An example of a rule that does not satisfy IIA is the following.
Suppose the cost of each agent is normalized by the range of values that this cost attains over the feasible set $\mathcal X$.
\[
C(\mathbf J(x)) = \sum_i \frac{J_i(x)}{\max_{x' \in \mathcal{X}} J_i(x') - \min_{x' \in \mathcal{X}} J_i(x')}.
\]
Let $\mathcal X = \{x, y, z\}$ with $\mathbf J(x)=(2,4)$, $\mathbf J(y)=(4,2)$ and $\mathbf J(z)=(3,3)$: the normalization ranges are both~$2$, so $C(\mathbf J(x)) = C(\mathbf J(y)) = 3$, a tie. Now change only the costs at~$z$ to $\mathbf J(z)=(1,8)$. The ranges become~$3$ and~$6$, giving $C(\mathbf J(x)) = \frac43$ and $C(\mathbf J(y)) = \frac53$, so the tie breaks in favor of~$x$, even though nothing about $x$ or $y$ has changed.

This non-locality is undesirable as it exposes the social ranking to manipulation: by altering how an irrelevant alternative is evaluated, one can change the ranking between two others. IIA prevents that.
The axiom can be relaxed by the designer, but only in a structured, deliberate, and interpretable way,
for example by specifying a benchmark point, as in the Nash social cost function discussed in the main text, or cost extrema, as in the Kalai-Smorodinsky solution \cite{Kalai1975}.

\sdbarfig{%
\includegraphics{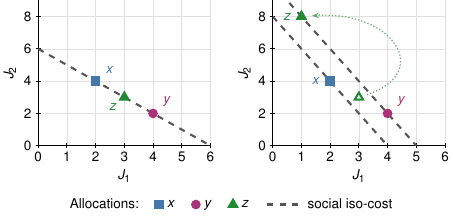}%
}{Changing the costs at an irrelevant dispatch~$z$ changes the normalization ranges, breaking the tie between $x$ and $y$. The IIA axiom prevents this dependence on unrelated alternatives.\label{sfig:iia}}

\subsection*{Without continuity: fragile rankings}
Continuity prohibits social cost rules whose values jump under small changes in the agents' evaluated costs. Without it, a strict social ranking between two outcomes can be reversed by an arbitrarily small perturbation.

Continuity is a core well-posedness property when social rankings are computed or embedded in numerical optimization procedures. In dynamic control settings, measurements fluctuate over time, are noisy, or arise within a feedback loop, and the optimal allocation may switch as a result (Figure~\ref{sfig:chattering}). Continuity does not prevent such switching near a tie, but it confines it to allocations that are almost equally good, so the welfare at stake is small. Without continuity, measurement noise can flip the decision between allocations that are far apart in social terms.

\sdbarfig{%
\includegraphics{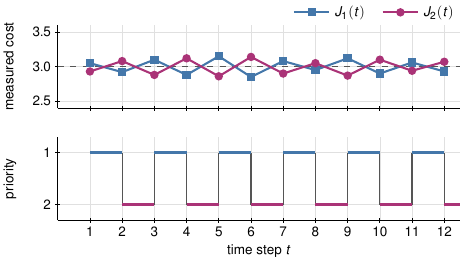}%
}{Noisy cost measurements around a tie make the optimal allocation switch. Under a continuous social cost function, the alternating allocations are almost equally good; without continuity, the same noise can flip between allocations with very different social cost.\label{sfig:chattering}}

\end{sidebar}

\begin{axiomp}{IIA}{Independence of Irrelevant Alternatives}
For any two outcomes $x, y \in \mathcal{X}$ and any two cost profiles $\mathbf{J}, \mathbf{J}'$ such that $J_i(x) = J'_i(x)$ and $J_i(y) = J'_i(y)$ for each~$i$, the social ranking between $x$ and $y$ must be the same under~$\mathbf{J}$ and~$\mathbf{J}'$.
\end{axiomp}

The ranking of two outcomes should depend only on how agents evaluate \emph{those} outcomes, not on costs at other outcomes. Without (IIA), the social ranking becomes vulnerable to manipulation: by adding, removing, or altering irrelevant alternatives, an agent (or the designer) can change the ranking between two outcomes without changing anything about how they are evaluated.

\begin{axiomp}{C}{Continuity}
For all $x, y \in \mathcal{X}$ and every sequence of cost profiles $\mathbf{J}^{\nu}$ converging pointwise to $\mathbf{J}$: if $x \succeq_{\mathbf{J}^{\nu}} y$ for all~$\nu$, then $x \succeq_{\mathbf{J}} y$.
\end{axiomp}

If outcome~$x$ is weakly preferred to~$y$ under every profile of a convergent sequence, it remains so in the limit: the social ranking is robust to small errors in the costs.

An additional property that is often desirable, but \emph{not} required for the welfarist approach, is Anonymity.

\begin{axiomp}{A}{Anonymity}
For any permutation $\pi: \mathcal{N} \to \mathcal{N}$: if $J'_i(x) = J_{\pi(i)}(x)$ for all $i$ and $x$, then $\succeq_{J} = \succeq_{J'}$.
\end{axiomp}

Anonymity encodes equal treatment, regardless of the index: if two agents swap their cost functions, the social ranking does not change.
It is natural in applications where all agents have equal standing but may \emph{not} be appropriate when agents have pre-determined priority levels.

\medskip

The following theorem shows that any SCFL satisfying the first three axioms admits a numerical representation by a single real-valued social cost function.

\begin{theorem}[Welfarism {\cite[Thm~3.7]{dAspremontGevers2002}}]
\label{thm:welfarism}
Let $\succeq_\mathbf{J}$ be a social preference relation on~$\mathcal{X}$, with $|\mathcal{X}| \geq 3$, defined by a SCFL $F$ for any profile $\mathbf{J} \in \mathcal{J}^n$.
If $\succeq_\mathbf{J}$ satisfies \emph{(P)}, \emph{(IIA)}, and \emph{(C)}, then there exists a continuous Social Cost Function
\[
C: \mathbb{R}^n \to \mathbb{R},
\]
such that for every profile $\mathbf{J}$ and any $x, y \in \mathcal{X}$,
\[
x \succeq_\mathbf{J} y \quad \Longleftrightarrow \quad C\big(J_1(x), \dots, J_n(x)\big) \leq C\big(J_1(y), \dots, J_n(y)\big).
\]
\end{theorem}

This result is the foundational step of the welfarist approach. Starting from a general SCFL, it shows that, under three natural axioms, the \emph{entire} social ranking induced by the SCFL can be captured by a function of the individual costs $\mathbf{J}(x)$. The problem of allocation then reduces to choosing and optimizing $C(\mathbf J(x))$ over the feasible set~$\mathcal{X}$. No additional information about the outcomes is needed beyond the individual cost vector $\mathbf J(x)$.

\begin{pullquote}
The Welfarism Theorem tells us that the social ranking can be represented by a \emph{social cost function} of individual costs, but it does not tell us \textit{which function}. That is determined by \emph{comparability assumptions}.
\end{pullquote}

The axioms guarantee that the social ranking is representable by a social cost function~$C$ of individual costs, but they do not pin down its form. Selection of one over the others now depends on how much the designer can compare costs \emph{across} agents, formalized by the notion of \emph{interpersonal comparability}.

%% file: sections/01_tutorial/04_comparability.tex
\subsection{Interpersonal Comparability}
\label{subsec:comparability}

The concept of interpersonal comparability addresses the following question: \emph{what kind of cross-agent cost comparisons are meaningful in a given application?}

Interpersonal comparability is formalized through \emph{invariance conditions}: 
formally, the social ranking must satisfy
\begin{equation*}%
    C(\mathbf J(x)) \leq C(\mathbf J(y)) \quad \Longleftrightarrow \quad C(\varphi \circ \mathbf J(x)) \leq C(\varphi \circ \mathbf J(y)),
\end{equation*}
for all $x, y \in \mathcal{X}$, every cost profile $\mathbf J$, and every $\varphi \in \Phi$, where $\Phi$ is a set of $n$-tuples $\varphi = (\varphi_1, \dots, \varphi_n)$ of increasing transformations and $\varphi \circ \mathbf J = (\varphi_1 \circ J_1, \dots, \varphi_n \circ J_n)$.

\begin{figure}
    \centering
    \includegraphics{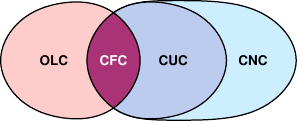}
    \caption{Nesting of comparability classes. Moving inward means the designer trusts \emph{more} information about how agents' costs relate to each other: CFC assumes full knowledge of cost scales and levels, while CNC requires only existence of a cardinal utility for each agent, without any comparisons being made. In practice, the choice of class is a statement about how well the designer knows, and can compare, the agents' true preferences.}
    \label{fig:comparability_venn}
\end{figure}

That is, if the designer only trusts certain features of the cost data (ordinal rankings, ratios, differences), then transformations that preserve those features should not alter the social decision.
By varying the set~$\Phi$, one obtains different comparability classes, each yielding a different admissible SCF family. We present four classes of transformations (see Figure~\ref{fig:comparability_venn}) and, for each class, we provide the admissible transformations and the resulting SCF.

\paragraph*{Ordinal Level Comparability (OLC)}

Under OLC, we assume only that costs can be \emph{ordinally} compared across agents: the meaningful statements can be ``agent~$i$ is worse off than agent~$j$,'' but not ``agent~$i$'s cost is twice as large.''
Formally, any common strictly increasing transformation of costs is admissible:
\[
\varphi_i^{\mathrm{OLC}}(t) = \varphi(t), \quad \varphi \text{ strictly increasing, same for all } i.
\]
By applying any such $\varphi$ to all costs simultaneously, only the ordinal ranking of costs across agents is preserved; no cardinal information (ratios, differences) survives.

Under (P), (IIA), (C), (A), and OLC invariance, the admissible SCFs are the positional rules $C(\mathbf J(x)) = J_{(k)}(x)$, the $k$-th largest cost \cite{Gevers1979}. One could thus minimize the cost of the best-off agent, or of the median one. Following Rawls \cite{Rawls1971}, the usual choice is to add an equity requirement \cite{Hammond1976}: a transfer from a better-off agent to a worse-off one, which leaves the ranking of the two unchanged and affects no one else, is never socially worse. Among the positional rules, only the worst-off position satisfies it, which yields
\begin{equation}\label{eq:OLC_SCF}
    C^{\mathrm{OLC}}(\mathbf J(x)) = \max_{i \in \mathcal{N}} J_i(x).
\end{equation}
This is the Rawlsian \emph{maximin} rule: the social cost equals the cost of the worst-off agent.

\emph{When is OLC appropriate?}
When only ordinal welfare information is available across agents, so that the designer can tell who is worse off under a given outcome, but not by how much. 
In other words, it must be possible to define equal treatment.

\begin{table*}[tb]
    \centering
    \renewcommand{\arraystretch}{1.1}
    \caption{Interpersonal comparability classes and their implications for the choice of social cost function.}
    \label{tab:comparability_summary}
    \begin{tabular}{@{}l l l l l@{}}
    \toprule
    \textbf{Class} & \textbf{Invariant transformations} & \textbf{Meaningful statements} & \textbf{Admissible SCF} & \textbf{Example: Power curtailment} \\
    \midrule
    OLC & $\varphi(J_i)$, common increasing & Ordinal ranking across agents & $\max_i J_i$ \emph{(maximin)} & Equal curtailment \\[3pt]
    CNC & $a_i J_i + b_i$, independent\ scales & Ratios of cost improvements & $-\prod_i [J_i(x^\circ)-J_i]^{c_i}$ \emph{(Nash)} & Proportionally fair curtailment \\[3pt]
    CUC & $a J_i + b_i$, common scale & Cost differences across agents & $\sum_i c_i J_i$ \emph{(utilitarian)} & Minimize total curtailment \\[3pt]
    CFC & $a J_i + b$, fully common & Levels \emph{and} differences & Mean $+ g$(deviations) & Efficiency--equity trade-off \\
    \bottomrule
    \end{tabular}
\end{table*}

\paragraph*{Cardinal Non-Comparability (CNC)}

Under CNC, each agent's cost function is meaningful on a cardinal scale. That means that we can say agent~$i$'s cost decreases by twice as much from~$x$ to~$y$ as from~$x$ to~$z$, but the scales \emph{cannot be compared across agents}.
Formally, any agent-specific positive affine transformation is admissible:
\[
\varphi_i^{\mathrm{CNC}}(t) = a_i t + b_i, \quad a_i > 0, \quad b_i \in \mathbb{R}.
\]
Under CNC, a well-known impossibility result \cite[Thm.~8*2]{Sen1970CCSW}, \cite{dAspremontGevers2002} shows that, for $|\mathcal X| \geq 3$, the only SCFs satisfying (P), (IIA), and (C) on a universal domain are dictatorial, $C(\mathbf J(x)) = J_k(x)$ for some agent~$k$; with Anonymity (A), none remains.
However, if we relax (IIA) to \emph{Partial Independence}, which allows the SCF to depend also on one fixed \emph{benchmark outcome}~$x^\circ$, the admissible SCF is the Nash bargaining solution \cite{Roberts1980, shilov2025welfare}
\begin{equation}\label{eq:CNC_SCF}
        C^{\mathrm{CNC}}(\mathbf J(x)) = -\prod_{i \in \mathcal{N}} \big[ J_i(x^\circ) - J_i(x) \big]^{c_i}, \quad c_i \geq 0,
\end{equation}
where at least one $c_i \neq 0$, and $x^\circ$ is a benchmark such that $J_i(x^\circ) > J_i(x)$ for all feasible~$x \neq x^\circ$ (a ``worst-case'' reference). With Anonymity (A), all exponents are required to be equal, hence positive.
Engineering applications usually have a benchmark ``worst-case'' reference: fallback policies, no or minimal service, or failure-mode performance. See \cite{MattBolognani2026} for an example of meaningful benchmarks in the energy domain.

The exponents $c_i$ can encode agents' relative entitlements when there are valid reasons to relax Anonymity (A), which would otherwise require them to be equal.
For example, they can represent different ownership or investment shares (in a simple profit-sharing problem with monetary payoffs, exponents proportional to investment yield equal gains per unit invested).
They can implement different access rights and priority levels. 
They can also account for the number of individuals represented by each agent \cite{Kalai1977nonsymmetric}.
Examples of applications in which the choice of these exponents is discussed include congestion control in data networks \cite{KellyMaullooTan1998}, climate policy \cite{Yu2017climate}, and water distribution \cite{Janjua2025water}.

\emph{When is CNC appropriate?}
When costs are cardinal for each agent (for example, monetary costs, measurable penalties) but the relative scaling across agents is unknown or unverifiable. Examples include electricity markets with unknown marginal costs, transportation systems with heterogeneous values of time, power demand response systems with private discomfort or flexibility costs, and network congestion control where data flows aggregate the demand of an unknown number of users.

\paragraph{Cardinal Unit Comparability (CUC)}

Under CUC, cost \emph{differences} are fully comparable across agents, ``one unit of cost for agent~$1$ is socially equivalent to one unit of cost for agent~$2$,'' but cost \emph{levels} may differ by arbitrary additive constants.
Formally:
\[
\varphi_i^{\mathrm{CUC}}(t) = a t + b_i, \quad a > 0, \quad b_i \in \mathbb{R}.
\]
Under (P), (IIA), (C), and CUC invariance, the admissible SCF is the weighted utilitarian form \cite{Roberts1980, dAspremontGevers2002}:
\begin{equation}\label{eq:CUC_SCF}
    C^{\mathrm{CUC}}(\mathbf J(x)) = \sum_{i \in \mathcal{N}} c_i J_i(x), \quad c_i \geq 0.
\end{equation}
where at least one $c_i \neq 0$. The weights $c_i$ encode the relative social importance assigned to the agents. Their choice can reflect the considerations discussed above for the Nash exponents, such as entitlements, access rights, priority levels, or the number of individuals represented by each agent.
Under Anonymity (A), they must be equal, giving the unweighted sum.

\emph{When is CUC appropriate?}
When the designer can place cost differences on a common scale across agents, even if the absolute levels of cost remain agent-specific. This is natural when agents face a common marginal unit of loss or benefit, but differ by additive offsets, for example because of different endowments, reservation utilities, or costs that lie outside the controlled part of the system.

\paragraph*{{Cardinal Full Comparability (CFC)}}

CFC is the strongest assumption: both cost levels and differences are fully comparable.
A single common affine transformation can be applied to all agents:
\[
\varphi_i^{\mathrm{CFC}}(t) = a t + b, \quad a > 0, \quad b \in \mathbb{R}, \quad \text{same for all } i.
\]
Under (P), (IIA), (C), and CFC invariance, all admissible SCFs have the form \cite{Roberts1980}
\begin{equation}\label{eq:CFC_SCF}
    C^{\mathrm{CFC}}(\mathbf J(x)) = \frac{1}{n}\sum_{i \in \mathcal{N}} J_i(x) + g\!\left( \begin{bmatrix} J_1(x) - \bar{J}(x) \\ \vdots \\ J_n(x) - \bar{J}(x) \end{bmatrix} \right),
\end{equation}
where $\bar{J}(x) = \frac{1}{n}\sum_i J_i(x)$ is the average cost and $g: \mathbb{R}^n \to \mathbb{R}$ is homogeneous of degree one\footnote{A function $g : \mathbb{R}^n \to \mathbb{R}$ is homogeneous of degree one if $g(\lambda z) = \lambda g(z)$ for every $z \in \mathbb{R}^n$ and every $\lambda > 0$.} and such that $C^{\mathrm{CFC}}$ is increasing in each~$J_i$. The first term is the utilitarian component; the second, through the choice of $g$ and of its weight $\gamma \in [0,1]$, is an inequality measure of the cost distribution. Two admissible examples are $g(z) = \gamma \max_i z_i$, which interpolates between the utilitarian ($\gamma=0$) and the maximin ($\gamma=1$) SCF \cite{BossertKamaga2020}, and the Gini mean difference $g(z) = \frac{\gamma}{2n^2}\sum_{i,j} |z_i - z_j|$, which yields the Gini social evaluation function $\bar J(x)\,(1+\gamma\, G(\mathbf J(x)))$.

\emph{When is CFC appropriate?}
Only when the designer can justifiably compare both cost levels and cost differences across agents on a common scale. That means that cost increments must be directly measurable in the same physical or monetary units, and it is possible to tell when two agents are treated equally (i.e., are incurring the same cost).

%% file: sections/01_tutorial/05_summary.tex
\subsection{Welfarist Design Procedure}
\label{subsec:comparability_summary}

The welfarist design procedure for multi-agent decision-making can now be stated as a simple recipe:
\begin{enumerate}
    \item \textbf{Identify the agents' preferences.} Define the feasible set $\mathcal{X}$ and the individual cost functions $J_i$.
    \item \textbf{Determine the level of comparability.} Ask: can cost \emph{differences} be compared across agents (CUC)? Can both cost \emph{levels} and \emph{differences} also be compared (CFC)? Can \emph{nothing} cardinal be compared across agents (CNC)? Are only ordinal rankings available (OLC)? The comparability level then restricts the social cost function to an admissible class (Table~\ref{tab:comparability_summary}).
    \item \textbf{Fix the SCF parameters.} Decide whether Anonymity (A) is retained or relaxed, and justify the weights $c_i$ in \eqref{eq:CUC_SCF} or the exponents $c_i$ in \eqref{eq:CNC_SCF} accordingly, for example through entitlements, access rights, or the number of individuals represented by each agent. Under CNC, select the benchmark $x^\circ$. Under CFC, select the inequality term $g$ and its weight $\gamma$ in \eqref{eq:CFC_SCF}.
    \item \textbf{Solve and report.} Optimize the resulting social cost function over $\mathcal{X}$, and state the adopted comparability level and the resulting SCF with its parameters as part of the design specification.
\end{enumerate}
We refer to \cite{shilov2025welfare} for some simple examples of how this procedure can be applied to decision problems in engineering systems: water distribution, transportation, and energy.

\input{sections/01_tutorial/06_curtailment_example_2}

%% file: sections/01_tutorial/06_curtailment_example_2.tex
\begin{sidebar}{Illustrative Example: Fair Power Generation Curtailment (cont'd)}\label{sidebar:fair-curtailment-2}

\sdbarinitial{L}et us revisit the curtailment problem under a welfarist lens, and see how the proposed design procedure applies.

\subsection*{Step~1: Identify the agents' preferences}
Suppose prosumer $i$ incurs a cost from curtailment according to
\[
    J_i(x_i) \;=\; \theta_i\,(\bar p_i - x_i).
\]
Parameters $\theta_i$ capture how costly one unit of curtailment is for prosumer~$i$.
They can depend, for example, on the scale of the agents (implying that the disutility reflects relative economic losses), the cost of generation, and the individual power remuneration contract.

\subsection*{Step~2: Determine the level of comparability}
 
 \textbf{Cardinal Unit Comparability (CUC):} We assume that cost differences can be compared across prosumers on a common scale, even if absolute cost levels may differ. 
In this case, it means that the parameters $\theta_i$ are known. 
If one unit of curtailment carries the same meaning for every agent ($\theta_i=\theta\ \forall i$), then CUC yields the utilitarian SCF. This may be plausible for a homogeneous fleet of identical rooftop systems facing the same tariff and operating conditions.

\noindent \textbf{Cardinal Non-Comparability (CNC):} If the $\theta_i$ are unknown and they are heterogeneous, then we should assume non-comparability. The Nash SCF is invariant under individual re-scalings of the agents' costs, which makes it appropriate when the operator cannot verify the $\theta_i$.

\noindent \textbf{Ordinal Level Comparability (OLC):} It is appropriate when the operator can rank prosumers by how adversely they are affected by curtailment, but cannot evaluate those losses on a cardinal scale. The design then focuses on the worst affected prosumer. 
In this example, with equal $\theta_i$, OLC yields the maximin SCF.

\subsection*{Step~3: Fix the SCF parameters}
\noindent No prosumer has a claim to priority, so Anonymity (A) is retained: the parameters $c_i$
are equal. Under CNC, the benchmark $x^\circ = 0$ (full curtailment) is a natural worst-case reference.

\subsection*{Step~4: Solve and report}
Optimize the chosen SCF over the feasible set~$\mathcal{X}$, and report the adopted comparability level together with the resulting SCF and its parameters. For example: ``We adopt CUC with equal weights because all prosumers have identical rooftop installations and face the same tariff.'' Or: ``We adopt CNC with equal exponents because prosumers differ in privately known sensitivities to curtailment, and none has priority access.'' This way, the same physical curtailment problem leads to different allocation rules depending on which welfare comparisons are justified. The three SCFs recover the three rules introduced in the first part of this example: minimum total curtailment (utilitarian), equal curtailment (maximin), and proportionally fair curtailment (Nash with the full-curtailment benchmark), see Figure~\ref{sfig:fair_curtailment_scf}.
More complex design choices can be made by introducing different interpretable benchmark allocations. See \cite{MattBolognani2026} for more sophisticated, but equally explainable, welfarist curtailment rules.

\sdbarfig{%
\includegraphics{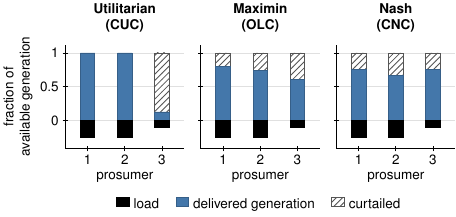}%
}{The same three allocations, now labeled by the social cost function and comparability class that produced them.\label{sfig:fair_curtailment_scf}}

\end{sidebar}

%% file: sections/01_tutorial/07_further_reading.tex
\subsection{Foundational Papers and Further Reading}

The aggregation of individual preferences into a collective decision is a standard question in social choice theory and welfare economics. A natural starting point is Arrow's impossibility theorem \cite{Arrow1951}. In a setting where individual preferences are ordinal and no interpersonal welfare comparisons are available, Arrow showed that no aggregation rule that maps every profile of individual orderings to a social ordering over at least three alternatives can satisfy unanimity (if every agent prefers one outcome to another then society must also prefer it), independence of irrelevant alternatives, and non-dictatorship at the same time. The result identifies a basic limitation of aggregation under a restricted informational basis.

Sen \cite{Sen1970CCSW, Sen1977} showed that this limitation changes once some interpersonal comparisons are admitted. If one can compare agents to a limited extent, for example by saying that agent~$i$ is worse off than agent~$j$, or that agent~$i$ gains more from moving from~$x$ to~$y$ than agent~$j$ does, then additional aggregation rules become admissible. In this sense, the form of the social ordering depends not only on the axioms imposed on the aggregation rule, but also on the information on cross-agent welfare comparisons.

Two well-known cases illustrate this. Harsanyi \cite{Harsanyi1955} derived the weighted utilitarian form from expected-utility rationality of individuals and society together with Pareto indifference, with interpersonal comparisons entering through the weights. Under much weaker informational assumptions, Rawls \cite{Rawls1971} proposed the \emph{maximin} criterion, which gives priority to the welfare of the worst off. Within Sen's framework, this corresponds to ordinal level comparability. Nash's bargaining solution \cite{Nash1950} represents an ``intermediate position''. More generally, Roberts \cite{Roberts1980} and d'Aspremont and Gevers \cite{dAspremontGevers2002} characterized the admissible aggregation rules associated with different degrees of interpersonal comparability. We recommend \cite{Roberts1980} as a starting point for the interested reader for social choice from an economics perspective, and our own paper \cite{shilov2025welfare} in the language of control.  

For further references, see \cite{FleurbaeyHammond2004} and \cite{BossertWeymark2004} for more general axiomatic characterizations of social welfare functionals and a general treatment of utility in social choice. See \cite{Hammond1991} for a survey of interpersonal comparisons of utility in welfare economics, and see \cite{BlackorbyBossertDonaldson2002} for the relation between utilitarianism and theories of distributive justice. For a related direction concerning bargaining and cooperative game theory, including Nash and Kalai--Smorodinsky cooperative solutions, and also Shapley's work on utility comparisons in games see \cite{Nash1950,Kalai1975,Shapley1969UtilityComparisons,Thomson2022}. 
For dynamic settings of social choice, as we shall discuss later, two lines of literature are relevant, namely social evaluation under risk and uncertainty \cite{MonginPivato2016} and intertemporal or inter-generational social choice \cite{BossertSuzumura2015,PivatoFleurbaey2024}. 

%% file: sections/01_tutorial/08_proxy.tex
\begingroup\raggedcolumns%
\begin{sidebar}{When Is Our Cost Model Good Enough?}\label{sidebar:proxy}
\setlength{\bigskipamount}{5pt}\setlength{\abovecaptionskip}{3pt}\setlength{\belowcaptionskip}{0pt}\setlength{\abovedisplayskip}{5pt}\setlength{\belowdisplayskip}{5pt}%
\sdbarinitial{I}n real-world applications, agents' preferences and true cost functions are usually not known to the designer. They may be withheld for privacy, since the data that reveal them are sensitive \cite{MolinaMarkham2010}, or they may be misreported when the allocation depends on the report \cite{Gibbard1973,Chremos2024MechanismDesign}.
Reliance on elicited subjective preferences may be problematic in regulated infrastructures, where operational decisions are often required to rely on objective, transparent, and verifiable criteria \cite{EU2019_944}. 

The designer therefore works with a \emph{proxy}, an observable map $\mathbf{s} = (s_1, \dots, s_n)$ standing in for the true but unobservable cost profile~$\mathbf J$.
Measured proxies cannot be misreported, and they can therefore be trusted.
Because they are a model of true preferences, it is natural to ask whether the allocation is then fair with respect to the agents' true costs, or only to the designer's model of them.
In the language of social choice, the question is whether ranking by $C(\mathbf s(x))$ agrees with ranking by $C(\mathbf J(x))$.

\subsection*{Consistency}

Let $\mathcal C_\Phi$ be the set of SCFs admissible under a comparability class~$\Phi$. The proxy~$\mathbf s$ is \emph{$C_\Phi$-consistent} on $D \subseteq \mathcal X$ if, for every $C \in \mathcal C_\Phi$ and all $x, y \in D$,
\[%
    C\bigl(\mathbf J(x)\bigr) \leq C\bigl(\mathbf J(y)\bigr)
    \;\Longleftrightarrow\;
    C\bigl(\mathbf s(x)\bigr) \leq C\bigl(\mathbf s(y)\bigr),
\]%
that is, the proxy yields the social ranking that the true costs would.
If $\mathbf s = \varphi \circ \mathbf J$ for some $\varphi \in \Phi$, consistency follows from invariance. 
On a sufficiently rich domain, for instance when the attainable cost vectors form a comprehensive set, the converse holds as well under CUC and CFC: $\mathbf s$ is consistent on $D$ only if $\mathbf s = \varphi \circ \mathbf J$ on $D$ for some $\varphi \in \Phi$. Under OLC, a common increasing $\varphi$ with $\mathbf s = \varphi \circ \mathbf J$ suffices.
Under CNC, consistency is a weaker requirement, which only needs to hold between the benchmark $x^\circ$ and the operating region: with cost savings $\Delta_i = J_i(x^\circ) - J_i(x)$ and proxy savings $\tilde\Delta_i = s_i(x^\circ) - s_i(x)$, the Nash SCFs \eqref{eq:CNC_SCF} are consistent if and only if $\tilde\Delta_i = \lambda_i\, \Delta_i^{\alpha}$ (with agent-specific $\lambda_i > 0$ and a common $\alpha > 0$).

\subsection*{Consistency as an information filter}

The comparability class is thus both a normative stance on which comparisons across agents are meaningful and an \emph{information filter} on which features of the cost model are trusted.
Consistency serves as a specification that the proxy needs to meet. If it does not hold, then the designer must choose a weaker comparability level, change the SCF (e.g., select a different benchmark), or improve the proxy.

\subsection*{Example: Power generation curtailment cost}

The linear cost $\theta_i(\bar p_i - x_i)$ of the power curtailment example is itself a proxy choice. 
The true cost for the prosumer depends on their own power demand: local generation first replaces the prosumer's own purchase of expensive grid power, then it is used to serve additional own demand in order of decreasing marginal utility, then it is exported to the grid, when further self-consumption is worth less than the financial gain (see figure). 
The grid operator does not know the exact shape of these cost functions, as they depend on agent-specific consumption preferences and demand elasticity, and on how the financial gain is valued by each prosumer (e.g., proportionally to the investment, which is unknown).

However, if CNC is assumed and the benchmark $x_i^\circ$ is set at the zero-export point, the proxy $s_i$ and the true cost $J_i$ are consistent on allocations with positive export (they satisfy $\tilde\Delta_i = \lambda_i\, \Delta_i^{\alpha}$ because they are equal, up to an agent-specific affine transformation, in that regime). Therefore, measuring exported power gives enough information to make a fair social choice. See \cite{MattBolognani2026} for further discussion on this specific application.

\sdbarfig{%
\includegraphics{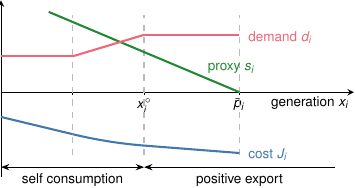}%
}{Schematic representation of the prosumer demand and the resulting cost (power purchase cost minus power sale and demand utility) as a function of the prosumer generation $x_i$, plotted together with the proxy $\theta_i(\bar p_i - x_i)$ used to guide the curtailment decision.}

\end{sidebar}

\endgroup

%% file: sections/02_dynamic_welfarism.tex
\begin{pullquote}
    Rather than certifying the welfarist nature of each decision, control engineers must guarantee that the closed-loop behavior of the socio-technical system is correctly aligned with the agents' preferences.
\end{pullquote}

When automating socio-technical systems, control engineers face specific challenges but also unique opportunities, compared to the classical setting of social choice that we reviewed in the short tutorial.

In standard social choice problems, the utility or cost incurred by an agent is a function of the outcome, which is entirely under the authority of the decision maker.
Social choice was mostly developed for static, one-shot allocation problems (voting, allocation of rights of use of a public common, etc.). 

In contrast, in engineering contexts, when social choice theory is applied, the decision maker affects the outcome in a more complex way (see Figure~\ref{fig:dynamic-social-choice}):
\begin{itemize}
    \item \textbf{Dynamics: } the outcome may correspond to the output of a dynamical system driven by the decision; 
    \item \textbf{Uncertainty: } the dynamics of the system may be partially unknown and stochastic;
    \item \textbf{Disturbances: } other time-varying and unmeasured factors also affect the system and thus the outcome.
\end{itemize}

Control engineers traditionally approach these challenges via \emph{feedback}: if the outcome can be measured, then its measurement can be used to drive adjustments of the decision variable towards the optimal decision and to reject the unmeasured disturbances.
For the purpose of social choice, this means that rather than certifying the welfarist nature of each decision, control engineers must guarantee that the closed-loop behavior of the socio-technical system is correctly aligned with the agents' preferences.

To illustrate how the design of a controller can be aligned to closed-loop welfarist specifications, we select three control methods that naturally allow incorporating a performance objective in their design: \emph{feedback optimization}, \emph{stochastic optimal control}, and \emph{model predictive control}.
In three short tutorials, we show what challenges arise when a welfarist objective (an SCF) is employed. 

\emph{Feedback optimization} can drive a stable dynamical system to a welfarist outcome, even in the presence of disturbances. %
We can only certify welfarism of the steady state, not of the transient of the plant.

In \emph{stochastic optimal control}, the designer needs to decide how the welfarist objective is aggregated over time, and how it is evaluated on a stochastic outcome. Different choices lead to different welfarist certificates.

In \emph{model predictive control}, special care is needed to derive closed-loop guarantees when welfarist objectives are used in a receding horizon optimization. It is unclear how to derive welfarist certificates in the presence of constraints and how to certify welfarism of the plant trajectories.

Each method is accompanied by an example of application for which it seems particularly well suited.
The selected applications highlight three crucial features that characterize welfarism in socio-technical systems:
\begin{itemize}
    \item \textbf{Limited scope:} each of these decisions has a well-defined scope, where a specific resource is allocated (or a loss is divided);
    \item \textbf{Repeated decisions:} the social choice is repeated frequently in an iterative fashion, and the welfarist specifications often apply to infinite-horizon, long-term, or average outcomes;
    \item \textbf{Feedback:} online measurements of the outcome are available, allowing the decision maker to employ data-driven policies, react to external factors, and compensate inequalities as they materialize.
\end{itemize}
Under these conditions, control engineers have a unique opportunity to bring social choice from theory to operational practice.

\subsection{Welfarist Feedback Optimization}
\input{sections/02_dynamic_welfarism/01_feedback_optimization/02_datanetworks}

\input{sections/02_dynamic_welfarism/01_feedback_optimization/01_feedback_optimization}

\subsection{Welfarist Stochastic Optimal Control}
\input{sections/02_dynamic_welfarism/02_markov_decison_processes/02_mobility_service_allocation}
\input{sections/02_dynamic_welfarism/02_markov_decison_processes/01_mdp}

\subsection{Welfarist Model Predictive Control}
\input{sections/02_dynamic_welfarism/03_model_predictive_control/02_energy_management}
\input{sections/02_dynamic_welfarism/03_model_predictive_control/01_mpc_new}

%% file: sections/02_dynamic_welfarism/01_feedback_optimization/02_datanetworks.tex
\begin{sidebar}{Example of Welfarist Feedback Optimization: Data Network Utility Maximization}

\sdbarinitial{T}he need for ``fair'' or ``equitable'' allocation of data network capacity has been observed since the very first studies of data congestion protocols as a control system \cite{chiu1989analysis,Low2002}, making it an early success story of welfarist control.
Both proportional \cite{KellyMaullooTan1998} and max-min \cite{mo2000fair} policies have been identified as valid candidates, to grant fair access to the shared infrastructure and to enhance the protocol's robustness against uncertain knowledge of the agents' needs and of the network parameters.

To illustrate these policies through the lens of dynamic welfarism, consider this simple example: a server is transmitting video streams to three agents via a wireless channel, and needs to decide the transmitting power $p_i$ towards each of them.
Let us assume a simple signal interference-noise model, so that the data rate that is delivered to each agent is 
\[
r_i = \frac{p_i w_i}{D_i} \quad D_i := 1 + \varepsilon \sum_{j\ne i} p_j,
\]
where $w_i$ is a time-varying fading parameter (representing channel attenuation, environmental interference, and background traffic)
and $\varepsilon$ is the intensity of the cross-channel interference.
The rates $r_i$ are the measurable outcome of the decision problem and are determined by both the decisions $p_i$ and the exogenous factors $w_i$. 
We do not have access to a model of the dynamic transient, and we assume it quickly converges to its steady state.
A proxy for the agents' utility (negative cost) is the framerate of the delivered videos
\[
f_i(r_i) = 30 \min\left\{\frac{r_i}{R_i},1\right\},
\]
where $R_i$ is the data rate needed to deliver 30 frames per second for the video stream $i$. 

Depending on the comparability assumption, we aim at minimizing different social cost functions: Rawlsian \eqref{eq:OLC_SCF}, Nash \eqref{eq:CNC_SCF}, or utilitarian \eqref{eq:CUC_SCF}.
The minimum of these functions can be tracked over time via \emph{feedback optimization}.
Namely,  we iteratively adjust the transmitting power by following a gradient update, where the gradient of the welfare cost (or subgradient, when nonsmooth) can be determined via chain-rule 
and is then projected to the budget constraints
\[
p_i \ge 0\ \forall i, \quad \sum_i p_i = 3.
\]

The closed-loop behavior of the controller shows that the choice of how to trade off agents' costs (i.e., the comparability level) drastically determines the real-time allocation of limited network capacity:
\begin{itemize}
    \item \emph{Utilitarian} -- The assumption of cardinal unit comparability means that we trust the proxy to represent the agents' utilities on the same scale, and it therefore makes sense to allocate the entire power budget to the agents that can use it most efficiently, and starve the others.
    \item \emph{Nash} -- For this specific model and proxy, and while no agent saturates, the Nash update rule is independent of the measured outcome and its equilibrium is equal power to the agents (\emph{equal allocations} rather than \emph{equal outcomes}). By construction, it guarantees proportionality: allocating more power to any agent would increase its utility less than it would decrease the others' utilities, in proportional terms.
    \item \emph{Max-min} -- As the only possible comparison between agents is ordinal, the allocation rule tries to achieve equal framerates, no matter how inefficient that can be.
\end{itemize}

\sdbarfig{%
\includegraphics{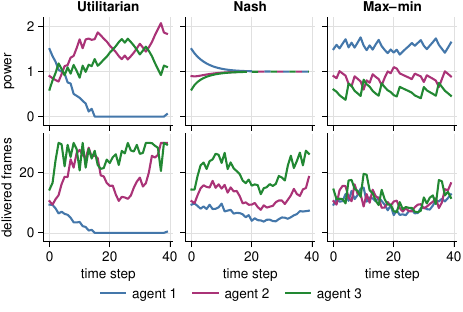}
}{Simulation of the decisions and outcomes when feedback optimization is employed with different SCFs. The fading parameter of the wireless channel changes in time and requires a time-varying allocation of transmitting power.}

\end{sidebar}

%% file: sections/02_dynamic_welfarism/01_feedback_optimization/01_feedback_optimization.tex
Consider the dynamic social choice problem of Figure~\ref{fig:dynamic-social-choice}. Assume that the dynamical system 
\[
\dot x = f(x, u, w)
\]
that maps the decisions $u$ to the 
outcomes $x$ (which determine the agents' costs $J_i(x)$) is stable and has fast transients.
$w$ represents unmeasured exogenous factors that can vary slowly over time.
In control theory terms, $u,x,w$ correspond to system input, output, and disturbance, respectively.

In this setting, we may aim at updating the decision $u$ based on feedback measurements of the outcome $x$ so that the system is driven to an optimal outcome $x^\star$ that is best aligned (in a welfarist sense) to the preferences of the agents.
We assume that the cost $\mathbf J(x)$ of the agents is available in functional form, together with an appropriate comparability level.
As discussed earlier, the comparability level guides the choice of the SCF $C(\mathbf J(x))$, which in turn allows us to define the steady-state optimization problem 
\begin{equation}
\begin{split}
    (u^\star, x^\star) \  = \qquad  \argmin_{u,x} & \quad C(\mathbf J(x)) \\
    \subjectto & \quad x = h(u, w)\\
    & \quad u \in \mathcal U,
\end{split}
\label{eq:ofo}
\end{equation}
where $\mathcal U$ is the feasible decision space (e.g., delimited by budget constraints on the decision), and $h(u,w)$ is the steady-state map of the system.

Feedback optimization \cite{Hauswirth2024,Colombino2020} allows one to design iterative policies for the choice of $u$ based on the measurements $x$ so that the closed-loop system converges to and tracks the solution of \eqref{eq:ofo}, without measuring $w$.
Typically, a time-scale argument is invoked to neglect the fast plant dynamics and to guarantee stability and convergence (see \cite{Hauswirth2021} for a rigorous presentation), although it is not always necessary \cite{Yousefi2025, Bianchi2025}.
These guarantees require smoothness (Lipschitz gradients) and typically convexity of the cost; these hold for weighted sums and can be checked or enforced (smoothing, staying away from the benchmark) for maximin and Nash.

For the design of these iterative updates, we can tap into the literature of nonlinear optimization and employ methods like projected gradient descent (although many other options are possible, including primal-dual iterations).
Let 
\begin{equation}
    \widetilde C(u) := C(\mathbf J(h(u, w))),
\label{eq:compositeOFOcost}
\end{equation}
which corresponds to the SCF evaluated on the decision, rather than on the outcome. 
Because $w$ is unknown, $\widetilde C(u)$ in general cannot be evaluated, but it is instrumental for the design procedure.
The projected gradient descent step takes the form
\begin{equation}
u^{t+1} = \Pi_{\mathcal U}\left[ 
    u^{t} - \alpha \nabla \widetilde C(u^{t})^\top
\right]
\label{eq:ofoupdate}
\end{equation}
where the gradient $\nabla \widetilde C$ can be computed via chain-rule differentiation on the composite function \eqref{eq:compositeOFOcost}:
\[
\nabla \widetilde C(u)  =  \nabla C(\mathbf J(h(u,w))) \nabla \mathbf J(h(u,w)) \nabla_u h(u, w)
\]
Notice that the differentiation requires knowing the sensitivity $\nabla_u h(u,w)$, that is, the marginal effect of the decision on the outcome\footnote{In some cases, this first-order information on the model is not available.
See \cite[Section 3.3]{Hauswirth2024} for a discussion of alternatives that rely on online sensitivity learning or derivative-free iterations (e.g., zeroth order optimization \cite{He2024}, extremum seeking \cite{Ariyur2003}).},
but does not require evaluating $h(u,w)$ numerically, as the outcome $x$ is measured.
In fact, assuming fast convergence of the system, we equivalently have
\[
\nabla \widetilde C(u)  = \nabla C(\mathbf J(x)) \nabla \mathbf J(x) \nabla_u h(u, w).
\]

To gain some intuition on the resulting iterative updates of the social choice $u$, we briefly compute the iterative update \eqref{eq:ofoupdate} for the different social cost functions that we reviewed earlier. 

\paragraph{Feedback utilitarian optimization}

In this case, $C(\mathbf J(x)) = \sum_{i=1}^n J_i(x)$,
therefore %
the update is
\[
u^{t+1} = \Pi_\mathcal{U} \left[ 
u^{t} - \alpha \nabla_u h(u, w)^\top \sum_{i=1}^n \nabla J_i(x^{t})^\top
\right].
\]
An interesting special case emerges when the utility $J_i$ of agent $i$ depends only on the elements $x_i$ of the outcome, and $x_i$ depends only on the element $u_i$ of the decision.
This is the case, for example, when the decision maker can allocate resources to agents individually, but the individual outcomes are also affected by external factors $w$, i.e., $x_i = h_i(u_i,w)$.
This yields the simplified update
\[
u^{t+1} = \Pi_\mathcal{U} \left[ 
u^{t} - \alpha 
\begin{bmatrix}
    \nabla_{u_1} h_1(u_1, w) \nabla J_1(x_1) \\
    \ldots \\
    \nabla_{u_n} h_n(u_n, w) \nabla J_n(x_n)
\end{bmatrix}
\right].
\]

Notice that the projection $\Pi_\mathcal{U}$ on the budget constraints means that the resource is moved from other agents toward the agents with the highest marginal utility sensitivity, starving  the agents that require more resources to achieve the same quality level (even for reasons beyond their control; see \cite{Zhan2024FairOFO} for a discussion in a practical setting).
This approach is efficient in theory, but the standard social choice results reviewed earlier tell us that it relies on a high comparability level and model accuracy, i.e., no uncertainty about the scale of the marginal utilities $\nabla J_i(x_i)$ and of the plant sensitivities $\nabla_{u_i} h_i(u_i, w)$. 

\paragraph{Feedback Nash optimization}

Consider, for simplicity, negative $J_i$ (i.e., positive utilities of the agents) and assume cardinal non-comparability with identical $c_i$ and benchmark $x^\circ$ such that $J_i(x^\circ)=0$ for all $i$. 
Then the social choice problem consists 
of minimizing the equivalent SCF 
\[
\widetilde C(J(x)) = -\sum_{i=1}^n \log (-J_i(x)).
\]
Then the gradient takes the form
\[
\nabla \widetilde C(u) = 
-
\begin{bmatrix}
    \frac{1}{J_1(x)} & \ldots & \frac{1}{J_n(x)}
\end{bmatrix}
\nabla J(x)
\nabla_u h(u,w).
\]

In the special case in which the cost of agent $i$ depends only on the element $x_i$ of the outcome, the gradient descent step becomes 
\[
u^{t+1} = \Pi_\mathcal{U} \left[ 
u^{t} + \alpha \nabla_u h(u,w)^\top
\begin{bmatrix}
    \frac{\nabla J_1(x_1)}{J_1(x_1)} \\ 
    \vdots \\ 
    \frac{\nabla J_n(x_n)}{J_n(x_n)}
\end{bmatrix}
\right],
\]
which confirms that the update is invariant to scaling of the individual cost functions.

\paragraph{Feedback Rawlsian optimization}

Consider the SCF
\[
C(\mathbf J(x)) = \max_i J_i(x)
\]
that derives from an assumption of ordinal level comparability between the individual costs $J_i(x)$.
The objective is nondifferentiable, therefore a subgradient descent method can be employed (notice that closed-loop asymptotic stability may be compromised, see discussion in \cite[Section III.C]{Hauswirth2021}).
The subgradient of $\widetilde C(u)$ is given by the convex hull of the directions
\[
\{ \nabla J_i(x) \nabla_u h(u,w) : i \in I_{\max}(x) \},
\]
where $I_{\max}(x)$ is the set of worst-case agents:
\[
I_{\max}(x) := \{i:  J_i(x) = \max_j J_j(x)\}.
\]

When the worst-case agent is unique and denoted by $i^\star$, the update takes the form
\[
u^{t+1} = \Pi_\mathcal{U} \left[ 
 u^{t} - \alpha \nabla_u h(u,w)^\top \nabla J_{i^\star}(x^{t})^\top
\right].
\]
The update transfers resources to the agent that is suffering the highest cost. 
Crucially, this assessment is based on the functional cost $J_i$ evaluated on the measured outcome $x$.
In other words, the closed-loop nature of the dynamic social choice implements a form of distributive justice, i.e., it aims at equality of outcomes and compensates for the external factors $w$.

\paragraph{Open challenges}

An important observation is that the guarantees of welfarism only apply to the steady state: the single iterative updates of the decision variables, in general, do not satisfy any welfarist specification, and no welfarist claims can be made on the transient of the system. No compensation for unfair transients is happening either.
While this is almost inconsequential when looking at the long-term behavior of the undisturbed system, it does persistently affect the dynamic social choice in case of time-varying $w$. 
The impact of this residual effect on agents may not be optimal with respect to the chosen SCF: different agents may be affected differently and may pay an ``unfair price'' for the time-varying disturbances.

Second, feedback optimization comes with robustness guarantees against model mismatch, specifically with respect to uncertainty about the plant steady-state sensitivities $\nabla_u h(u,w)$. 
In a multi-agent setting, we need to certify how the model mismatch affects the different agents individually. The earlier utilitarian example showed that the choice of comparability level should be dictated not only by how much we trust the ratio of scales between different agents' costs, but also by how accurately we know the plant sensitivities.

Finally, in case of cardinal non-comparability, the definition of a valid SCF requires the designer to choose a benchmark outcome $x^\circ$ (an attainable worst case).
The online determination of such a reference point, in the context of a time-varying $w$, is a challenge \emph{per se}: if the only available model information is the steady-state sensitivities, it is difficult to predict what the outcome would be at a different location, let alone finding the worst-case attainable outcome $x^\circ$ that can be attained via a feasible decision $u^\circ \in \mathcal U$.

%% file: sections/02_dynamic_welfarism/02_markov_decison_processes/02_mobility_service_allocation.tex
\begin{sidebar}{Example of Welfarist Stochastic Optimal Control: Public Transit}

\newcommand{\tc}[0]{\mathrm{c}}
\newcommand{\ts}[0]{\mathrm{s}}

\sdbarinitial{T}ransportation systems such as public transit and ride-hailing must repeatedly allocate scarce service capacity under uncertainty. These operational decisions have distributional consequences: serving one zone better often worsens service in another~\cite{Ferguson2012incorporating}. 
We study a spare-bus allocation problem as an example of welfarist stochastic optimal control.

Consider a transit operator providing service to two zones: a dense urban core (\emph{city}) and a \emph{suburban} area. Each day $t$, suburb demand $d_\ts^t$ is fixed while city demand $d_\tc^t$ switches stochastically between a moderate level (\textbf{M}) and a high level (\textbf{H}) due to tourism and city events, as shown in Figure~\ref{sfig:bus:mdp}.

\sdbarfig{%
\includegraphics{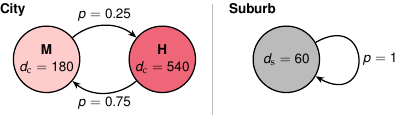}%
}{Demand transition probabilities for the two zones. 
\label{sfig:bus:mdp}}

The operator has 13 buses and provides $(n_\tc,n_\ts)=(6,3)$ buses to the city and suburb for a baseline service. 4 spares cover demand uncertainty and potential maintenance needs. Daily allocations $u_\tc^t, u_\ts^t$ specify adjustments to these baseline levels. After each day, each operating bus independently requires one day of maintenance, with probability 0.02 in the city and 0.01 in the suburb. The total maintenance counts $b_\tc^t, b_\ts^t$ therefore follow binomial distributions.
Buses undergoing maintenance are unavailable the next day, hence the allocations must satisfy $u_\tc^{t+1} + u_\ts^{t+1} \leq 4 - b_\tc^t - b_\ts^t$.
Negative allocations represent reducing the baseline service. We assume that no buses are under maintenance at time $t=0$.

Assuming random demand arrival, the expected waiting time for each traveler is $\frac{\tau_i}{2(n_i+u_i^t)}$ at zone $i \in \{\tc,\ts\}$, where $\tau_\tc = \tau_\ts = 60$ min is the bus cycle time. 
We consider the \emph{total reduced waiting time} as a measure of zone-level utility:
\begin{equation}
    V_i^t(u_i^t) = d_i^t \left(\frac{\tau_i}{2 n_i} - \frac{\tau_i}{2(n_i + u_i^t)}\right), \ t = 0, \ldots, T-1.
\end{equation}
The utilities are treated as fully comparable (CFC). 
We then translate the obligation for providing equitable services via a maximin criterion ($\min_{i \in \{\tc,\ts\}}V_i^t$) to protect whichever zone gains less from spare allocation.

\subsection{Comparison of different welfare criteria}
We then decide when to apply the chosen SCF across time and uncertainty.
Do we care about how the zones are treated each day or over the time horizon? Should we provide equitable service on average, or for each realized operation scenario? The four welfare criteria in Table~\ref{tab:mdp:criteria} reflect different answers to these questions. We compare them over $T=7$ days.

Figure~\ref{sfig:bus:total_benefit} reports expected cumulated utility over the horizon per zone. Under utilitarianism, all criteria coincide. More spare buses are allocated to the city, as the city admits a higher marginal utility due to the higher demand level. Maximin narrows the gap between the two zones. In particular, CES and SCE nearly equalize expected cumulated utility. SEC and ECS evaluate ex-post welfare, leading to smaller variances. Since SEC focuses on stage-wise welfare, it is Pareto dominated by others when comparing accumulated utilities.

\sdbarfig{%
\includegraphics{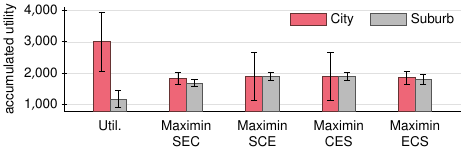}%
}{Accumulated utility of each zone over 7 days under utilitarian and maximin with different welfare criteria. Results are generated from Monte Carlo simulation with 1,000 runs. Error bars indicate one standard deviation. \label{sfig:bus:total_benefit}}

Figure~\ref{sfig:bus:control_input} and Figure~\ref{sfig:bus:equality_gap} compare the criteria under maximin across two different demand scenarios: a typical week with $\approx 25\%$ high-demand days, and an all-high week. 

SEC enforces equality on a daily basis. On high-demand days it allocates more to the suburb to match the benefits gained by the city.
CES and SCE plan based on the probability of high/moderate demand. They allocate more buses to the city on high-demand days and compensate the suburb on moderate days. They achieve near equality in a typical week but leave large inequality in extreme scenarios.
Differently, ECS tracks the past and compensates along the realized sequence, substantially reducing end-of-horizon inequality in both scenarios.

\sdbarfig{%
\includegraphics{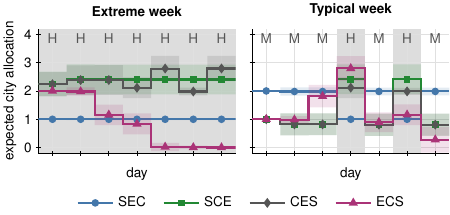}%
}{Expected city allocation under four criteria for two scenarios. All remaining available spares are allocated to the suburb. Gray shaded columns represent high city demand. Shaded bands around curves indicate one standard deviation.\label{sfig:bus:control_input}}

\sdbarfig{%
\includegraphics{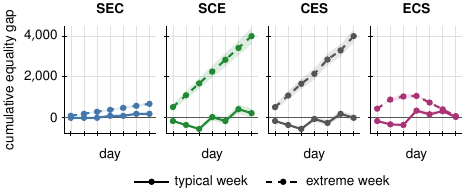}%
}{Cumulative equality gap ($\sum_{t'=0}^{t} V_\tc^{t'}-V_\ts^{t'}$) under four criteria for two demand scenarios. Shaded bands indicate one standard deviation.
\label{sfig:bus:equality_gap}}

\end{sidebar}

%% file: sections/02_dynamic_welfarism/02_markov_decison_processes/01_mdp.tex
Uncertainty enters welfarist control problems in multiple ways: system states evolve probabilistically~\cite{zhang2014Fairness,wen2021Algorithms,Mandal2023Socially,ju2023Achieving,fan2023Welfare}, agents' preferences may stochastically shift over time~\cite{parkes2013Dynamic,kulkarni2020Social,banerjee2022Online,huang2024Online}, and resource supply or agent participation may fluctuate~\cite{aleksandrov2015Online,kash2014no,sinclair2023Sequential}. In this section, we study welfarist stochastic optimal control in such multi-agent stochastic systems.

In single-objective stochastic optimal control problems, control designers aggregate costs along two dimensions: across \emph{time} (typically via summation $\sum_t$) and over \emph{uncertainty} (typically via expectation $\mathbb{E}[\cdot]$ if risk-neutral). 
In a multi-agent setting, the SCF $C(\cdot)$ enters the problem as the third aggregation operator, applied across agents. Since $C(\cdot)$ is generally nonlinear and therefore not interchangeable with the other two operators, 
it is important to carefully decide the order in which the three operators are composed. 
Each distinct ordering of the aggregators defines a \emph{welfare criterion} that represents different normative positions and leads to different outcomes. 
Table~\ref{tab:mdp:criteria} describes all of them, organized by two features: the welfare scope (whether $C(\cdot)$ is applied at each stage or to the sequence), and ex-ante or ex-post (whether $C(\cdot)$ evaluates expected or realized costs). Some criteria coincide, as expectation and summation commute. Clearly, they all coincide when a utilitarian SCF is employed, as it is linear.

SCE and SEC assess welfare at each stage, while CES and ECS assess welfare from cumulative costs and therefore allow compensation across stages.
SCE and CES evaluate the welfare of average outcomes, and hence a realized outcome might incur very undesirable welfare. SEC and ECS instead evaluate welfare for each realized outcome before averaging. With a nonlinear SCF, this can make them sensitive to outcome variability that is not captured by average outcomes alone.

The following example illustrates how these criteria can rank control policies quite differently, providing intuition for their interpretation.

\begin{table*}[t]
    \centering
    \renewcommand{\arraystretch}{1.1}
    \caption{Four welfare criteria for welfarist stochastic optimal control.}
    
    \medskip
    
    \begin{tabular}{ccccc}
        \toprule
        \textbf{Criterion acronym*} & \textbf{Mathematical form} & \textbf{Welfare scope} & \textbf{Ex-ante/Ex-post} & \textbf{Rationale} \\
        \midrule
        SCE & $\sum_{t} C(\mathbb{E}\left[\mathbf{J}(x^t,u^t)\right])$ & At every time & Ex-ante & Welfare of average stage outcome \\
        SEC & $\sum_{t}\mathbb{E}\left[C(\mathbf{J}(x^t,u^t))\right]$ & At every time & Ex-post & Average welfare of stage outcome \\
        CES & $C(\mathbb{E}\left[\sum_{t} \mathbf{J}(x^t,u^t)\right])$ & Over time & Ex-ante & Welfare of the average sequence \\
        ECS & $\mathbb{E}\left[C(\sum_{t} \mathbf{J}(x^t,u^t))\right]$ & Over time & Ex-post & Average welfare of the sequence \\
        \bottomrule 
        \multicolumn{5}{l}{\scriptsize * The acronyms are mnemonics for the order of the operations: \textbf{S}um, social \textbf{C}ost function, \textbf{E}xpectation.}
    \end{tabular}
    \label{tab:mdp:criteria}
\end{table*}

\paragraph{Example: two-stage allocation}
    Consider assigning a task to two agents, twice. At each stage $t=1,2$, the controller can either assign the task to one agent, or split it evenly. In the first case, an agent incurs a cost of $10$ and the other $0$; in the second case they both incur a cost of $8$. 
    
\begin{figure}[tb]
    \setlength{\fboxsep}{0.2pt}
    \renewcommand{\arraystretch}{1.3}
    \newcommand{\textbfcol}[1]{
        {\fbox{\strut \,\textbf{#1}\,}}%
    }
    \centering
    \begin{tikzpicture}
    \node (tbl) {
        \begin{tabular}{@{}ccccc@{}}
             & \includegraphics[scale=0.4]{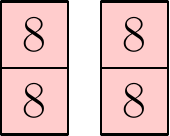} 
             & \includegraphics[scale=0.4]{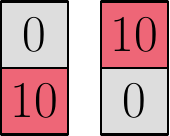}
             & \includegraphics[scale=0.4]{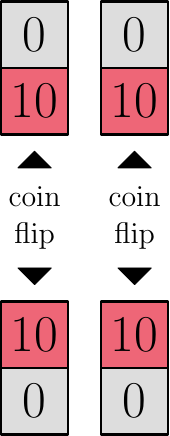}
             & \includegraphics[scale=0.4]{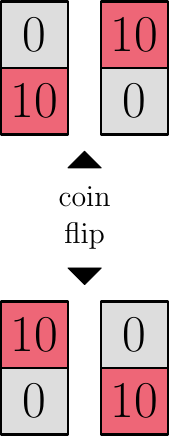} \\
             & determ. & determ. & stage-stoch. & seq.-stoch.\\
             \midrule
             SCE & 16 & 20 & \textbfcol{10} & \textbfcol{10} \\
             SEC & \textbfcol{16} & 20 & 20 & 20 \\
             CES & 16 & \textbfcol{10} & \textbfcol{10} & \textbfcol{10} \\
             ECS & 16 & \textbfcol{10} & 15 & \textbfcol{10} \\
             \bottomrule
        \end{tabular}
    };
    \node[anchor=north west] at ($(tbl.north west)+(+10mm,+4mm)$) {
        \includegraphics[scale=0.4]{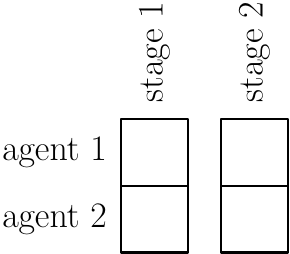}
    };
    \end{tikzpicture}
    
    \caption{Each column corresponds to a possible sequence of task assignments in the two-stage allocation example. The table lists the welfare costs computed via the criteria listed on the left. The first two sequences are deterministic. In the third one, the assignment is randomized at each stage. In the fourth one, the randomization is done over two possible assignment sequences.}
    \label{fig:coinflips}
\end{figure}

    Let us take a Rawlsian stance and aim at minimizing the largest cost (\emph{minimax}).
    As summarized in Table~\ref{tab:mdp:criteria}, there are four possible ways in which the SCF $C(\cdot) = \max_i (\cdot)$ can be employed to obtain a social ordering of the possible sequences of task assignments.
    In Figure~\ref{fig:coinflips} we use them to evaluate four possible decisions:
    \begin{itemize}
        \item the task is split every time;
        \item the task is assigned first to agent 1, then to agent 2;
        \item at every stage, the task is assigned with a coin flip;
        \item a coin flip decides whether agent 1 is assigned the task first and then agent 2, or vice versa.
    \end{itemize}

    The four criteria yield very different social orderings of these policies.

    SCE optimizes the welfare of the average stage outcome. It's an ex-ante notion of fairness and there is no reparation over time. Both randomized solutions are equally good, even if they produce uneven costs at every stage and possibly also over the two stages.

    SEC optimizes the average welfare of each stage. There is no reparation, but it considers the welfare of individual realizations. The best solution is to split at all times.

    CES optimizes the welfare of the average sequence. The two randomized policies are as good as a deterministic policy that assigns the task to one agent and then the other. 

    ECS optimizes the average welfare of the sequence. It prefers policies (stochastic or deterministic) whose realizations treat agents equally \emph{over time}. It guarantees that no agent incurs a total cost larger than 10.
    
\paragraph{Computation of optimal welfarist policies}
There are many settings and approaches for modeling and studying stochastic control problems. Markov decision processes (MDPs) are one of them. As a tutorial, for the rest of this section, we restrict our attention to MDPs and show how to compute optimal welfarist policies.

Consider an MDP $\mathcal{M} = (\mathcal{X}, \mathcal{U}, P, \mathbf{J}, \rho)$, where $\mathcal{X}$ is the state space, $\mathcal{U}$ is the input space, $P_t(x' \mid x, u)$ is the transition probability at time step $t$, $J_i(x,u)$ is the stage cost incurred by agent $i$, and $\rho(x)$ is the initial state distribution. 

Given full knowledge of the MDP $\mathcal{M}$, we can formulate each welfare criterion as a convex program over occupancy measures whenever $C$ is convex~\cite{altman2021constrained}. We focus on a \emph{finite-horizon} setting with horizon $T$, while we refer the readers to \cite[Chapter 3-4]{altman2021constrained} for extensions to infinite-horizon settings with average or discounted costs.

For SEC, applying $C(\cdot)$ at each stage transforms the problem into a standard single-objective MDP. The stage social cost $C(\mathbf{J}(x^t,u^t))$ is a fixed scalar for every $(x^t,u^t)$ pair.
Given a policy $\pi$, the SEC objective equals
\begin{equation}\label{eq:mdp:ESC_occupancy_measure}
\begin{aligned}
    \sum_t \mathbb{E}\Big[
     C\big(\mathbf{J}(x^t,u^t)\big) \Big| \pi, \rho
    \Big] =
    \sum_t \sum_{x,u} \sigma^t_\pi(x,u) C\big(\mathbf{J}(x,u)\big),
\end{aligned}
\end{equation}
where $\sigma^t_\pi(x,u)$ is the stage-wise state-input occupancy measure, i.e., the probability of visiting $(x,u)$ at stage $t$ under $\pi$.
The optimal occupancy measure $\sigma^*$ solves
\begin{subequations}\label{eq:mdp:ESC}
\begin{align}
    \textbf{SEC:} \quad \min_{\sigma} \quad & \sum_t \sum_{x,u} \sigma^t(x,u) C\left(\mathbf{J}(x,u)\right) \label{eq:mdp:ESC:obj} \\
    \text{s.t.} \quad & \sum_u{\sigma^t(x,u)} = \sum_{x',u'} P_t(x | x',u') \sigma^{t-1}(x',u'), \nonumber \\ 
    & \hspace{2cm} \forall x \in \mathcal{X}, t \in [1,T-1] \label{eq:mdp:ESC:occ_measure_cstr1}  \\
    & \sum_u \sigma^0(x,u) = \rho(x), \ \forall x \in \mathcal{X} \label{eq:mdp:ESC:occ_measure_cstr2} \\
    &\sigma^t(x,u) \geq 0, \ \forall x \in \mathcal{X}, u \in \mathcal{U}, t \in [0,T-1].\label{eq:mdp:ESC:occ_measure_cstr3}
\end{align}
\end{subequations}
Constraints \eqref{eq:mdp:ESC:occ_measure_cstr1}-\eqref{eq:mdp:ESC:occ_measure_cstr3} enforce legal occupancy measures with respect to the transition probability $P$. 
And the corresponding optimal policy $\pi^{t,*}(u | x)$ for all $x \in \mathcal{X}, u \in \mathcal{U}$ can be reconstructed by
\begin{equation}\label{eq:mdp:opt_policy}
    \pi^{t,*}(u | x) = \begin{cases}
        \frac{\sigma^{t,*}(x,u)}{\sum_{u'} \sigma^{t,*}(x,u')}, & \sum_{u'} \sigma^{t,*}(x,u') > 0 \\
        \frac{1}{|\mathcal{U}|}, & \text{otherwise.}
    \end{cases}
\end{equation}

Because $C(\mathbf{J}(x,u))$ is precomputed, \eqref{eq:mdp:ESC} is \emph{always} a linear program (LP), regardless of which SCF is chosen.

Similarly, one can compute the optimal occupancy measure for SCE and CES with modified objective
\begin{align*}
    & \textbf{SCE:} \ \min_{\sigma} \ \sum_t C \Big(\sum_{x,u} \sigma^t(x,u) \mathbf{J}(x,u)\Big) 
    \ \text{s.t.} \ \eqref{eq:mdp:ESC:occ_measure_cstr1}-\eqref{eq:mdp:ESC:occ_measure_cstr3}  %
    \\
    & \textbf{CES:} \ \min_{\sigma} \ C\Big(\sum_t\sum_{x,u} \sigma^t(x,u) \mathbf{J}(x,u)\Big) 
    \ \text{s.t.} \ \eqref{eq:mdp:ESC:occ_measure_cstr1}-\eqref{eq:mdp:ESC:occ_measure_cstr3}  %
\end{align*}
and recover the optimal policy via \eqref{eq:mdp:opt_policy}.
For Utilitarian and Rawlsian, these two problems admit LPs with proper reformulation~\cite{zhang2014Fairness}, and can be solved efficiently. For Nash welfare, if the negative weighted geometric mean $C(\mathbf J)=-\left(\prod_i \Delta_i^{c_i}\right)^{1/\sum_i c_i}$ is adopted, where $\Delta_i = J_i^\circ - J_i$ are the savings with respect to the benchmark costs $J_i^\circ$ and $c_i > 0$, the SCF is convex and both SCE and CES admit convex problems.

\paragraph{Connections to dynamic programming}
Dynamic programming (DP) is commonly adopted for optimal control problems in MDPs. 
We dedicate a note to build the connection between DP and the method based on occupancy measure described above.

Standard DP requires a \emph{recursively separable} objective and the Markov property to induce a Bellman recursion. Then \emph{at each stage}, the optimal control input can be chosen independently \emph{for each state}.
SEC satisfies such requirements and its occupancy-measure LP~\eqref{eq:mdp:ESC} is equivalent to DP~\cite{altman2021constrained}. 
In contrast, SCE and CES generally do not admit a standard Bellman recursion.
CES objective is not recursively separable due to the generally nonlinear $C(\cdot)$ applied after the temporal summation. 
For SCE, although recursive separability is preserved, $C(\cdot)$ after the expectation introduces coupling across states at each stage. Hence, the optimal control cannot be determined independently for each state. Backward induction can still be carried out for SCE by optimizing jointly over all states.

\paragraph{Non-Markovian policies for ECS}

For SEC, $C(\cdot)$ simply transforms the stage cost, reducing the problem to a standard single-objective MDP. 
And for SCE and CES, $C(\cdot)$ is applied to \emph{expected} stage cost or \emph{expected} cumulative costs. 
They can be fully expressed using occupancy measures.
By \emph{the dominance of Markovian policies}~\cite[Chapter 2.4]{altman2021constrained}, Markovian policies $\pi^{t}(u^t | x^t)$ are sufficient to achieve any feasible occupancy measure. Hence, all these three criteria can be optimized by a time-indexed Markovian policy.

ECS is structurally different. Its objective 
evaluates welfare over the realized sequences. The ECS value hence depends on the \emph{joint} distribution of the sequence rather than the occupancy measure (marginal distribution) alone. To ensure that the welfare of every realized sequence is tracked, the controller may require a non-Markovian policy $\pi^t(u^t | x^0, u^0, x^1, u^1, \ldots, x^t)$ that makes decisions based on history. In the two-stage allocation example, ECS conditions the second-stage decision on the first-stage realization, compensating the previously disadvantaged agent.

In fact, rather than tracking the entire history, it suffices to augment the state with the accumulated cost $z^t_i = \sum_{t'<t} J_i(x^{t'},u^{t'})$. Define $\tilde{x}^t = (x^t, z^t)$ and modify the cost functions
\begin{equation}
    \widetilde{J}_i(\tilde{x}^t,u^t) =
        z^{t}_i + J_i(x^t,u^t).
    \label{eq:integrator}
\end{equation}
Given a policy $\pi$, the ECS objective can be expressed by 
\begin{equation*}
\begin{aligned}
    & \mathbb{E}\Big[
        C\big( \sum_t\mathbf{J}(x^t,u^t)\big) \Big| \pi, \rho
    \Big] \\ 
    = \ &\mathbb{E}\Big[
        C\big(\widetilde{\mathbf{J}}(\tilde{x}^{T-1},u^{T-1})\big) \Big| \pi, \rho
    \Big]
    = \sum_{\tilde{x},u} \sigma^{T-1}_\pi(\tilde{x},u) C\big(\mathbf{\widetilde{J}}(\tilde{x},u)\big).
\end{aligned}
\end{equation*}
This augmented MDP can then be solved by LP or DP, yielding a Markovian policy $\pi^{t,*}(u^t | \tilde{x}^t)$.

\paragraph{Open challenges}
Aggregation over time and uncertainty may demand more than just summation $\sum_t$ and expectation $\mathbb{E}[\cdot]$. 

First, social goals may call for different intertemporal aggregation: A planner minimizing the worst stage cost would use $\max_t$ rather than $\sum_t$; and a requirement of invariance to time-specific positive affine transformations (e.g., when there is a lack of intertemporal comparability) would suggest Nash product $\prod_t$~\cite{baqaee2010intertemporal}. On top of that, infinite-horizon settings may call for discounting future welfare. Properties of these objectives~\cite{Abate2008probabilistic}, with and without temporal discounting, are worth studying.

Second, for non-Markovian policies, integrating the past as in equation \eqref{eq:integrator} might not be satisfying. A society that forgets distant past outcomes would prefer a discounted or rolling-window aggregation (e.g., how a resource was allocated 10 years ago should not affect current decisions). The integration of such considerations can require different designs of state augmentation~\cite{alamdari2024Remembering,kumar2026Past}. 

Third, the social planner or the agents might dislike the risk-neutral $\mathbb{E}[\cdot]$. They may prefer risk measures such as CVaR~\cite{rockafellar2000optimization}. 
Meanwhile, as described before, ECS already shows some sensitivity to risk. How it relates to explicit risk-aware criteria remains unknown.

While all four criteria treat agents symmetrically, in practice, agents may have different temporal preferences or risk tolerances. A richer welfare language is needed to incorporate such heterogeneous preferences.

Last but not least, reinforcement learning is applied when the environment is unknown. Standard learning approaches like Q-learning 
do not directly generalize to all welfare criteria since they might not admit Bellman recursion. Online occupancy measure estimation~\cite{Mandal2023Socially,ju2023Achieving} and state augmentation~\cite{alamdari2024Remembering} or accumulated-reward-dependent policies~\cite{fan2023Welfare} are hence adopted to either bypass or restore Bellman recursion. Designing efficient and reliable learning algorithms in these welfarist settings, especially when the dimensionality becomes large, is an open research direction.

%% file: sections/02_dynamic_welfarism/03_model_predictive_control/02_energy_management.tex
\begin{sidebar}{Example of Welfarist Model Predictive Control: Energy Management}

\sdbarinitial{T}he green transition requires residential demand to align more closely with production peaks of renewables, essentially redistributing who can draw energy from the grid, in what quantity and when, especially during moments of scarcity when renewables are low or grids are near their power limits. Energy is essential for survival and participation in modern society and MPC is standard for control in energy management, as accurate dynamic grid models are available and hard constraints are necessary to prevent brownouts. 

Consider a set of households $\mathcal{N}$, each owning a battery with state of charge $q_i^t$. At each time step, household $i\in \mathcal{N}$ consumes energy $e_i^t$ and can charge or discharge its battery at rate $s_i^t$, incurring a cost
\begin{align*}
    J_i(u,x_i) &= \underbrace{(s_i^t+ e_i^t)\Big(\rho_{i,1}^t + \rho_{i,2}^t\sum_{j\in\mathcal{N}} (s_j^t+ e_j^t) \Big)}_{\text{energy cost}} + \underbrace{\rho_{i,3}^t (\zeta_i^t)^2}_{\text{shift discomfort}} \\&+  \underbrace{\rho_{i,4}^t (q_i^t)^2}_{\text{battery degradation}} 
\end{align*}

\startsidebarfig
\begin{wrapfigure}{l}{0.45\columnwidth}
  \centering
  \includegraphics[width=0.45\columnwidth]{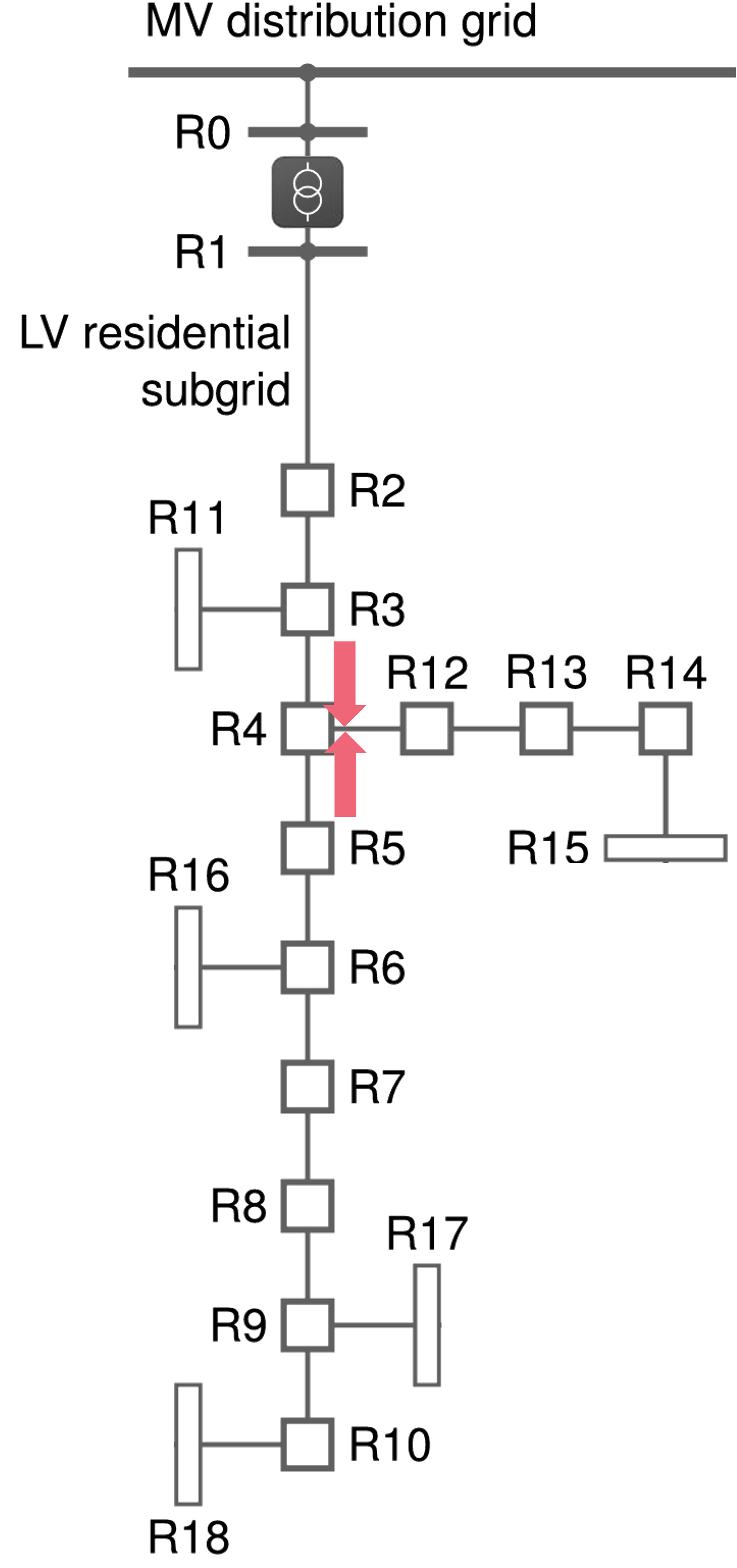}
  \caption{Simplified CIGRE grid model.}\label{fig:CIGREgridModel}
\end{wrapfigure}
\stopsidebarfig

where $s_i^t + e_i^t$ is the energy household $i$ purchases from the grid to satisfy its demand, charged at a base price $\rho_{i,1}$ plus a linear price $\rho_{i,2}$ scaling with aggregate consumption $\sum_{j\in\mathcal{N}} (s_j^t + e_j^t)$.  The parameters $\rho_{i, 3}, \rho_{i, 4}$ weigh their personal preference over comfort and battery degradation with respect to the energy cost. The shifting discomfort $\zeta_i^{t+1} = \zeta_i^{t} + (\tilde{e}_i^t - e_i^t)$ tracks how much consumers have shifted from their preferred consumption profile $\tilde{e}_i^t $. Each household's battery is following standard charging dynamics 
$q_i^{t+1} = \alpha_i \,q_i^{t} + \beta_i s_i^t$ with efficiency factors $\alpha_i, \beta_i$ and constraints $0\leq q^t_i\leq \bar q_i$. For simplicity, we impose only radial-lossless line power limits
\begin{align*}
   \bar{p}_i^t \geq \sum_{j \succeq i} \left( s_j^t + e_j^t \right) \quad \forall i, \forall t
\end{align*}
which constrain the per-line power flow at each time step. We employ a simplified CIGRE model (Figure~\ref{fig:CIGREgridModel}).

\subsection{Open-loop comparison across SCFs}

If we consider a ``Welfarism over time'' approach, i.e., $C\Big(\sum_{t=0}^{T-1} \mathbf{J}(x^t,u^t)\Big)$, the optimal open-loop solution computed by the MPC controller may expose one agent to high cost at the beginning and compensate that at a later time. See, for instance, the maximin solution in Figure~\ref{fig:LoadStoragComparison}: under the maximin SCF, no grid energy is delivered to consumer 5 in the first steps of the open-loop trajectory. However, when only the first step is applied in a receding-horizon manner, this may end up with that consumer never getting energy. Achieving Welfarism in closed-loop would require certification through theoretical guarantees.

\sdbarfig{%
\includegraphics{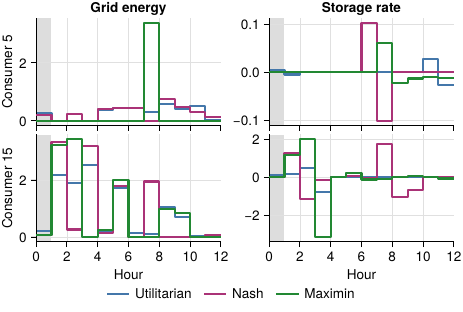}
}{Open-loop prediction for three different SCFs. Maximin minimizes the largest absolute cost here.\label{fig:LoadStoragComparison}}

\subsection{Cost redistribution under constraints }

Figure~\ref{fig:CostRedistribution} shows how total consumer cost changes when line capacity is reduced for consumers 12, 13, and 15 (pink arrow in Figure~\ref{fig:CIGREgridModel}). Consumer 15 has the largest load behind the constrained line, so it shifts the most energy. It also has a high discomfort weight and thus bears the largest cost increase. In the open-loop prediction, Nash and maximin spread the burden across all consumers, whereas under utilitarianism the affected consumers bear nearly all of it. This does not transfer to closed loop, where maximin burdens consumer 15 even more than utilitarianism. In receding-horizon control the SCF is applied to the predicted costs over each horizon, but its properties do not necessarily transfer to the realized trajectory, as maximin shows. Further, closed loop, total cost exceeds the utilitarian one by 4\% under Nash and by 13\% percent under maximin. %

\sdbarfig{%
\includegraphics{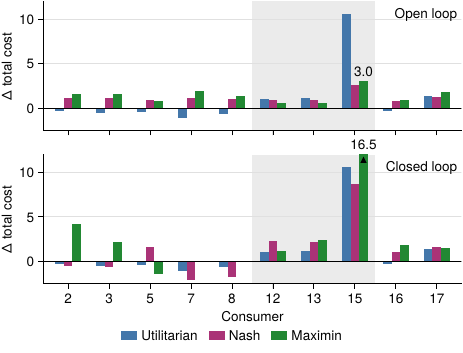}%
}{Change in cost when line capacity is reduced at hours 1--6 for consumers 12, 13, and 15 (pink arrow, Fig.~\ref{fig:CIGREgridModel}). Maximin minimizes the largest cost relative to no management.\label{fig:CostRedistribution}} 

\end{sidebar}

%% file: sections/02_dynamic_welfarism/03_model_predictive_control/01_mpc_new.tex
In many dynamic socio-technical systems, the designer aims at
\begin{itemize}
    \item maximizing welfare over an infinite time horizon;
    \item compensating for model mismatch, noise, and disturbances; 
    \item avoiding constraint violations, which would lead to depletion of the resource (groundwater), damage to the infrastructure (power grid), or service degradation (communication networks).
\end{itemize}
Oftentimes, a model of the natural or engineered dynamics (electrical, hydrological, scheduling, market interactions) is available, and it allows one to take welfarist decisions now while accounting for their future consequences on the system and on the agents.
An obvious control strategy candidate for such an implementation is model predictive control (MPC).

\paragraph*{Related works on equity and fairness in MPC}

The MPC community has previously dealt with fairness and equity considerations. For instance, \cite{villa2025fair} proposes a Fair-MPC framework that combines efficiency (per-agent distance from target), equality (Hoover index on per-agent control effort), and equity (Hoover index on per-agent error term). 
Fair-MPC has also been applied to energy systems, as in \cite{sun2024fairness}, which aims to achieve fairness across operating conditions by pre-processing the training data used in a data-driven prediction model inside an MPC. \cite{jiao2022online} combines an MPC for microgrid dispatch, which sets the total charging-station power, with a second stage that allocates this power using maximin among EVs, while \cite{moshahedi2023alpha} combines $\alpha$-fairness and MPC to perform perimeter control and balance traffic flows between neighborhoods ``fairly''. In what follows, we present a comprehensive framework applicable across domains, based on the dynamic social choice tools summarized in the previous section, to incorporate welfare considerations into MPC.

\paragraph*{\textbf{Welfarist objective design for receding-horizon control}}

MPC is an optimization-based receding-horizon control technique, and therefore, its central ingredient is the careful design of the objective function of the finite-horizon open-loop control problem that is solved at every iteration.

Namely, the designer needs to decide an objective, function of the $T$-long sequences of inputs and outputs, for the program
\begin{subequations}
\label{eq:MPCPerAgent}
\begin{align}
\label{eq:RunningCost}
\displaystyle \min_{x,\, u}  \quad  & \text{objective}\\
\textrm{s.t.} \quad &  x^{t+1} =  f(x^t, u^t)  \hspace{1.5em}  t \in [0, T-1] \label{eq:Constr1}\\
&x^t \in \mathcal{X}, \;  u^t \in \mathcal{U} ,  \hspace{1.25em} t \in [0, T-1] \label{eq:Constr2}\\ 
&x^0 = \mathbf{x}, \label{eq:Constr3}
\end{align}
\end{subequations}
where $\mathbf{x}$ is the current condition of the system, which could encode information about previous ``treatment'' or experience of individuals (e.g., discomfort, service quality, wait times).

\medskip

There are three ways to embed welfarism in \eqref{eq:MPCPerAgent}.

\medskip

\noindent\textbf{Tracking a welfarist steady state.}
The optimal steady state (in a welfarist sense) of the system can be determined as the solution of
\begin{align}
\label{eq:WelfaristSteadystate}
(\bar x, \bar u) =
\argmin_{u\in \mathcal{U} ,\, x \in \mathcal{X}}  &\; C( \mathbf{J}(x, u))\quad 
\textrm{s.t.} \quad  x =  f(x, u).
\end{align}
The objective of the MPC program can be designed to encode the tracking error with respect to this steady state:
\begin{equation}
    \label{eq:trackingerror}
    \sum_{t=0}^{T-1} \sum_{i\in\mathcal N}
    \|x_i^t - \bar x_i\|^2 + \|u_i^t - \bar u_i\|^2.
\end{equation}
Other than the specific way in which the steady state has been selected, this formulation amounts to a standard MPC problem.
Notice that despite the separability of the objective over time, the decisions at different time steps are coupled by the dynamics of the system: MPC aims at an optimal transient towards the steady state.

The tracking error is evaluated in the space of outcomes and decisions, not in the space of agents' costs. 
It is possible to replace the form \eqref{eq:trackingerror} with a different aggregation of the tracking errors of each agent, potentially with the same aggregations that we have reviewed before (sum of errors, Nash, maximin).

\medskip

\noindent \textbf{Welfarism at every time.}
By choosing the objective
\begin{equation}
\sum_{t=0}^{T-1}C\big(\mathbf{J}(x^t,u^t)\big),
\label{eq:welfarismateverytime}
\end{equation}
the designer ensures that the sum over the horizon of the stage-wise social cost is minimized. For Nash, the product form \eqref{eq:CNC_SCF} and its logarithmic reformulation are equivalent at a single time but can lead to different trajectories when summed over time.

This approach carries some degree of flexibility to adapt the strategy to the plant dynamics: it may be convenient to accept a higher social cost at one time step to achieve lower social cost in the future, or operate the system outside of a steady state.

This flexibility over time, however, does not extend to the individual agent. In other words, it does not consider the cost that each agent incurs over the horizon, but rather at every time.
While restrictive, this may be perceived as ``more fair'' and more explainable (e.g., why are my lights out but my neighbor's are on?). The overall nonlinear cost function remains time-separable, which allows one to apply theoretical results from economic MPC. This approach is adopted in \cite{moshahedi2023alpha, villa2025fair}.

\medskip

\noindent\textbf{Welfarism over time.}
Agents might prefer losing out more in one hour if, over the course of the day, it means they have increased their own welfare over the prediction horizon. 
This approach is implemented via an objective function that aggregates the trajectory costs of each agent:
\begin{equation}
\label{eq:welfarismovertime}
    C \bigg(\sum_{t=0}^{T-1} \mathbf{J}(x^t,u^t)\bigg).
\end{equation}
The extra degree of freedom allows increasing welfare at the expense of explainability, as the resulting receding-horizon trajectories may be counter-intuitive.
We showcase this phenomenon in the example of welfarist MPC for energy management, where the effect of ``shifting welfare'' over time is particularly prominent in the maximin implementation: one consumer receives no energy in the first time step and is compensated later.

Standard MPC proof techniques~\cite{gruene2017nonlinear} rely on time separability, which is lost in this form.
Thus, deriving theoretical guarantees is an open challenge.

\paragraph*{\textbf{Guarantees for Welfarist MPC}} 

The use of a welfarist objective in the open-loop MPC program does not automatically ensure welfarist allocations in closed-loop and in infinite horizon.
In the following, we discuss two important theoretical guarantees which come from economic and nonlinear MPC in the context of dynamic welfarism.

\medskip

\noindent \textbf{Stability of the welfarist steady state.}
Stability in welfarist MPC certifies that the closed-loop system converges to the welfarist steady state, so the system operates predictably instead of drifting or cycling. The standard tool employed to prove it in economic or nonlinear MPC is dissipativity~\cite{gruene2017nonlinear,faulwasser2018economic}. For the objective \eqref{eq:welfarismateverytime} that ensures \emph{Welfarism at every time}, and with the terminal constraint $x^T = \bar x$ added to \eqref{eq:MPCPerAgent}, the welfarist closed loop is asymptotically stable at~\eqref{eq:WelfaristSteadystate}~\cite[Thm.~2]{angeli2012average} provided the system is strictly dissipative with respect to the supply rate $s(x,u) := C(\mathbf{J}(x,u)) - C(\mathbf{J}(\bar x,\bar u))$, i.e., there exists a storage function $\lambda: \mathbb{X} \to \mathbb{R}$ and a positive definite $\rho$ such that%
\[
    \lambda(f(x,u)) - \lambda(x) \leq s(x,u) - \rho(\|x-\bar x\|).
\]
Strict dissipativity is well studied for stage costs that are strictly convex in $(x,u)$ with linear dynamics \cite{diehl2011lyapunov, damm2014exponential}. There exist convex reformulations of Nash for which this should apply, while maximin stage costs are in general not strictly convex.

In case of a quadratic objective, as is the case of the \emph{Tracking} approach \eqref{eq:trackingerror}, stability may be certifiable also for other aggregation functions.
Some results from log-barrier MPC could potentially transfer to Nash~\cite{wills2004barrier, feller2017relaxed}.
Showing dissipativity for maximin remains, to the best of our knowledge, an open problem. However, there are other techniques to prove stability in min-max MPC~\cite{raimondo2009min, limon2006input}, which however were developed in the context of robust MPC; it is not clear how or if such results transfer to the welfarist MPC setting.

We summarize stability results and point to relevant literature in Table~\ref{tab:welfareStabilityGuarantees}.

\begin{table}[tb]
\caption{MPC closed-loop stability guarantees (\emph{Tracking of Welfarist steady state} and \emph{Welfarism at every time}).}
\label{tab:welfareStabilityGuarantees}
\centering
\renewcommand{\arraystretch}{1.1}
\begin{tabular}{@{}lccc@{}}
\toprule
 & \textbf{Maximin} & \textbf{Nash} & \textbf{Utilitarian} \\ 
\midrule
{Quadratic objective} 
  & $\checkmark$~\cite{rawlings2017model}$^{*}$
  & $\sim$~\cite{faulwasser2018economic}
  & $\checkmark$~\cite{rawlings2017model}$^{*}$ \\ 
{General $J_i$} 
  & x
  & x
  & $\sim$~\cite{faulwasser2018economic} \\ 
  \bottomrule
\end{tabular}\\
\medskip
\footnotesize{$^{*}$ After shifting to the welfarist steady state, with terminal cost and terminal set satisfying \cite[Assumption~2.14]{rawlings2017model}.}\\
\footnotesize{$\sim$  Approaches from the cited paper can potentially be used or dissipativity can be shown to hold.}
\end{table}

\medskip

\noindent \textbf{Averaged social welfare.}
In receding horizon control, particularly with highly nonlinear objectives and dynamics, it may not be optimal to operate the system at the welfarist steady state~\eqref{eq:WelfaristSteadystate}. Instead, a time-varying allocation, e.g., a periodic cycle or limit cycle, can yield higher overall welfare \cite[Sec.~VI]{angeli2012average}. This phenomenon was formalized in economic MPC through the notion of averaged performance~\cite{angeli2012average}.
In the setting \emph{Welfarism at every time}, for the MPC with terminal constraint $x^T = \bar x$, it means that the averaged social welfare satisfies~\cite[Thm.~1]{angeli2012average}
\begin{align}\label{eq:AveragedWelfare}
      \limsup_{K\to \infty} \frac{\sum_{k= 0}^{K-1} C(\mathbf{J}(x^k,\kappa(x^k)))}{K}\leq C(\mathbf{J}(\bar x, \bar u))
\end{align}
where $k$ is now the time index for the closed-loop trajectory and $\kappa(x^k)$ is the welfarist MPC feedback law at the measured system state $x^k$. Intuitively, \eqref{eq:AveragedWelfare} certifies that the closed-loop trajectory, which may be periodic or follow a limit cycle, performs at least as well (in a welfarist sense) as operating the system at the welfarist steady state~\eqref{eq:WelfaristSteadystate}.

\paragraph*{\textbf{Open challenges}} 

As illustrated in the energy management example, the interplay between constraints and the choice of SCF strongly influences outcomes, particularly when constraints affect only a subset of the agents. 
This empirical observation motivates a closer theoretical study of how the dual variables in the Lagrangian are affected by the gradients of the specific SCF (which differ considerably, for instance, between Nash and maximin).
It is an open research question whether, in a constrained setting, the resulting closed-loop trajectories minimize the SCF. Nash and normalized maximin require a benchmark cost which may become infeasible under constraints. The choice of a feasible benchmark which is in line with the axioms and guarantees is still open.

A largely open topic is the analysis of the case that is not separable in time, i.e., the \emph{Welfarism over time} objective \eqref{eq:welfarismovertime}.
This setting has the attractive property of compensating utilities and losses over the prediction horizon. Yet, because the objective is non-separable in time, standard dissipativity and turnpike techniques of economic MPC do not apply.

%% file: sections/03_outlook_conclusion.tex
In this tutorial paper, we presented a comprehensive Welfarist design pipeline for control that promises to provide a principled foundation for fair and effective control. Some of its ingredients such as comparing and aggregating cost functions are shared with an emergent literature on fairness-oriented control methods. Our approach provides a rigorous theoretical underpinning based on strong foundations from the social choice literature. 
We demonstrate that social cost functions as admissible via Welfarism can be successfully used in common control methods such as online feedback optimization, stochastic optimal control, and MPC to tackle dynamic allocations, under uncertainty, while considering predictions and constraints. 

While those social results had traditionally been used in political philosophy and welfare economics as foundations of social contract theory, theories of redistribution and taxation, etc., their use in control contexts is direct and manifold, characterized by the following properties that draw specific boundaries for their use in contrast to the case of providing very fundamental social contract foundations.
Applications and control decisions are %
\begin{itemize}
    \item limited in scope (e.g., ``distributing energy to a fixed set of houses today'', rather than ``achieving a fair distribution of income in the entire society'');
    \item performed via faster online iterations (algorithms can be adapted and updated at any time, whereas policy changes slowly);
    \item compensated through feedback (to correct exogenous time-varying factors);
    \item repeated, and characterized by a dynamic nature that reduces the risk of any single bad allocation (past inequalities can be corrected over time);
    \item with a receding horizon which offers a tractable way to approximate infinite horizon policies (planning optimally for the infinite horizon future of an entire society, a problem considered in political philosophy, is clearly an intractable problem);
    \item intended for socio-technical systems that change rapidly, so a multi-generation perspective is not adequate or necessary.
\end{itemize}

Evidently, the foundations presented in this paper--as pointed out in the open challenges in each section--are just the start. Many control applications will require further development in order to fully translate and certify the axiomatic guarantees of social choice within the complex and interconnected control architectures that characterize control applications more generally. The following avenues for the development of such foundations appear particularly relevant to us:
\begin{itemize}
    \item \emph{Ex-ante vs.\ ex-post}: how to employ the perspective of feedforward vs.\ feedback control to design ex-ante fair decisions and certify/correct ex-post over time; is fairness an output regulation problem?
    \item \emph{Robustness and learning}: how to make axiomatic guarantees robust to different classes of stochasticity and uncertainty, how to learn the cost function model of agents online while controlling; is it possible to learn welfarist policies?
    \item \emph{Trajectory-wise optimization}: how do we extend computational methods in control (Markovian policies, receding-horizon optimization) to certifiably solve Welfarist decision problems over trajectories; can social welfare over time admit a Bellman principle? 
\end{itemize} 
In this paper, we express our view that once Welfarist control is further developed to address such real-world features, we will be able to make social choice applicable to a wide range of applications. The Welfarist framework that we envision promises to attain higher levels of transparency, fairness, and explainability, which will improve social acceptance of automation of critical infrastructure and automated allocation of essential resources.